\documentclass[review,12pt]{elsarticle}
\usepackage[margin=1in]{geometry}

\usepackage{type1cm}
\usepackage{makeidx}
\usepackage{graphicx}
\usepackage{multicol}
\usepackage[bottom]{footmisc}
\usepackage{amsmath,amssymb,amscd}
\usepackage{enumitem}
\usepackage[
colorlinks=false,
citecolor=black,
urlcolor=black,
linktocpage,
pagebackref
]{hyperref}
\usepackage{comment}
\usepackage{pgfplots}
\usepackage{algorithmic}
\usepackage{color}

\usepackage{listings}
\definecolor{mGreen}{rgb}{0,0.6,0}
\definecolor{mGray}{rgb}{0.5,0.5,0.5}
\definecolor{mPurple}{rgb}{0.58,0,0.82}
\definecolor{backgroundColour}{rgb}{0.95,0.95,0.92}
\definecolor{code}{rgb}{0.7, 0, 0.4}
\newcommand{\code}[1]{\texttt{\small\color{code} #1}}

\usepackage[ruled,vlined,linesnumbered]{algorithm2e}

\definecolor{darkgreen}{rgb}{0.0, 0.5, 0.0}

\SetKwRepeat{Do}{do}{while}

\def\bx{\mathbf{x}}   
\def\cmb{{\omega}}   
\def\lim{{\xi}}    

\def\ra{$r-$adaptivity}

\def\shsz{\emph{shape}+\emph{size}}

\def\bchi{\boldsymbol{\chi}}
\def\bw{\bar{w}}
\def\epsy{\varepsilon_y}
\def\epsz{\varepsilon_z}

\def\oneD{\mathrm{1D}}
\def\calO{\mathcal{O}}

\journal{Journal}

\begin{document}

\begin{frontmatter}

\title{Accelerating High-Order Mesh Optimization
       Using Finite Element Partial Assembly on GPUs }

\author{Jean-Sylvain Camier \fnref{llnl}}
\author{Veselin Dobrev \fnref{llnl}}
\author{Patrick Knupp \fnref{dihedral}}
\author{Tzanio Kolev \fnref{llnl}}
\author{Ketan Mittal \fnref{llnl}}
\author{Robert Rieben \fnref{llnl}}
\author{Vladimir Tomov \fnref{llnl}}
\fntext[llnl]
{Lawrence Livermore National Laboratory, 7000 East Avenue, Livermore, CA 94550}
\fntext[dihedral]
{Dihedral LLC, Bozeman, MT 59715}
\tnotetext[l_title]
{Performed under the auspices of the U.S. Department of Energy under
Contract DE-AC52-07NA27344 (LLNL-JRNL-835500)}

\address{}
\begin{abstract}
In this paper we present a new GPU-oriented mesh optimization method based on
high-order finite elements.
Our approach relies on node movement with fixed topology, through the
Target-Matrix Optimization Paradigm (TMOP) and uses a global nonlinear solve
over the whole computational mesh, i.e., all mesh nodes are moved together.
A key property of the method is that the mesh optimization process is recast in
terms of finite element operations, which allows us to utilize recent advances
in the field of GPU-accelerated high-order finite element algorithms.
For example, we reduce data motion by using tensor factorization and matrix-free
methods, which have superior performance characteristics compared to traditional
full finite element matrix assembly and offer advantages for GPU-based HPC
hardware.
We describe the major mathematical components of the method along with their
efficient GPU-oriented implementation.
In addition, we propose an easily reproducible mesh optimization test that can
serve as a performance benchmark for the mesh optimization community.
\end{abstract}

\begin{keyword}
mesh optimization \sep GPUs \sep performance benchmark
\sep finite elements \sep curved meshes \sep matrix-free methods
\end{keyword}
\end{frontmatter}


\section{Introduction}
\label{sec_intro}

The rise of heterogeneous architectures, such as general-purpose GPUs, has
motivated a rethinking of the algorithms used in large-scale,
high-performance simulations \cite{CEED2021}.
The new computing systems favor algorithms that expose ultra fine-grain
parallelism and maximize the ratio of floating point operations to
energy-intensive data movement.
As more applications are porting their codes to GPU systems,
meshing algorithms can become a bottleneck unless they can be
formulated in a balanced, data-parallel way.
Some of the existing mesh optimization methods are not a good match for the
new architectures because they rely on geometric operations and local topology
changes that introduce branching and load imbalance between parallel threads.

In this paper we present a computationally efficient GPU-capable
mesh optimization method based on high-order finite elements (FE).
Starting with our existing mesh optimization FE framework \cite{TMOP2019},
we reformulate all FE kernels using the so-called partial assembly (PA)
technique, and implement the resulting formulation through a
general abstraction that allows both CPU and GPU execution.
These modifications improve the method's computational performance
without affecting the result of the mesh optimization process.
Partial assembly is a technique introduced in the pioneering work of
Deville, Fischer, and Mund \cite{dfm02}, where the authors demonstrate that
by using a tensor product of 1D basis functions, fully assembled operators do
not need to be stored anymore and the computational cost
associated with the storage, assembly,
and evaluation of partially assembled operators reduces to
$\calO(p^{d})$, $\calO(p^{d})$, and $\calO(p^{d+1})$ per element,
respectively, where $p$
is the polynomial order of the FE space and $d$ is the space dimension.
The traditional FE approach of pre-assembling and storing sparse matrices,
on the other hand, leads to much worse computational costs per element, namely,
$\calO(p^{2d})$ for storage, $\calO(p^{3d})$ for assembly,
and $\calO(p^{2d})$ for evaluation.
Obtaining the above PA complexities, however, requires that the finite element
basis functions are tensor products of 1D basis functions, e.g.,
quadrilaterals in 2D and hexahedra in 3D.
Partial assembly has become even more relevant in recent years
\cite{bello2020matrix, franco2020high, kronbichler2019fast}
owing to its efficient use of GPU-based architectures, which are desirable for
arithmetically intensive applications that do not require a large amount of
data to be moved between the CPU and GPU \cite{CEED2021}.

The presented algorithms are developed in the context of the Target-Matrix
Optimization Paradigm (TMOP) for high-order meshes \cite{TMOP2019, TMOP2020}
and its implementation in the MFEM finite element library \cite{MFEM2020}.
TMOP minimizes a functional that depends on each element's current and
target (desired) geometric parameters: element {\em aspect-ratio}, {\em size},
{\em skew}, and {\em orientation}, which allows us to optimize and adapt the
mesh to improve the accuracy and computational cost of FEM
calculations (see, e.g., Section 4.1 of \cite{TMOP2021}).
The TMOP-based methodology is purely algebraic,
extends to both simplices and hexahedra/quadrilaterals of any order,
and supports nonconforming isotropic and anisotropic refinements in 2D and 3D.
Similar methods in the literature 
include Laplacian smoothing, where each node is typically moved as a linear
function of the positions of its neighbors \cite{vollmer1999improved,
field1988laplacian, taubin2001linear}, optimization-based smoothing, where a
functional based on elements' geometrical parameters is minimized
\cite{Knupp2012, Gargallo2015, mittal2019mesh,
      Greene2017, Peiro2018, aparicio2018defining},
and equidistribution with respect to an appropriate
metric tensor \cite{Huang1994, Huang2010, an2021moving}, amongst others
\cite{wallwork2020anisotropic, zint2020discrete, coulaud2016very}.

A survey of the literature shows that despite the introduction of recent
mesh optimization strategies on GPU-based architectures, the notion of
partial assembly is not present in the field of mesh optimization.
This is likely because most existing methods are either developed for
low-order meshes \cite{d2013cpu, Zint2018}
or use a localized approach (such as Laplacian smoothing or
optimization-based smoothing with a sequential patch-by-patch approach)
\cite{Zint2018, eichstadt2018accelerating, mei2014generic, shaffer2014simple}.
In contrast to other approaches, the variational-based TMOP methods are
well-suited to GPU acceleration, as all of the operations can be recast
in the form of finite element computations, allowing us to take
advantage of the significant GPU advances in this area.
In particular, matrix-free algorithms and partial assembly of nonlinear forms,
such as the global TMOP functional, can lead to orders-of-magnitude
reduction in the runtime of high- order applications compared to
traditional matrix-based algorithms \cite{CEED2020}.

The remainder of the paper is organized as follows. Section 2 gives
an overview of preliminaries such as high-order mesh representation and
the TMOP-based mesh adaptation framework, that are essential for understanding
our method. In Section 3, we describe the fundamentals of partial assembly for
high-order FEM operators. Here, we also describe our methodology for partial
assembly of TMOP-based mesh optimization methods.
In Section 4, we present several numerical
examples and benchmarks demonstrating the improvement in computational
efficiency due to use of partial assembly on GPU-based
architectures in comparison to CPU-based computations.
Finally, we close with directions for future work in Section 5.


\section{Preliminaries}
\label{sec_prelim}

This section provides a basic description of the TMOP-based \ra\ algorithm
that is the starting point of our work. We only focus on the aspects
that are relevant to the description of partial assembly of TMOP on GPUs.
See \cite{TMOP2019, TMOP2020} for additional details.


\subsection{Discrete Mesh Representation}
\label{subsec_mesh}

In our finite element based framework, the domain $\Omega \subset \mathbb{R}^d$
is discretized as a union of curved mesh elements of order $p$.  To obtain a
discrete representation of these elements, we select a set of scalar basis
functions $\{ \bw_i \}_{i=1}^{N_p}$, on the reference element $\bar{E}$.
In the case of tensor-product elements (quadrilaterals in 2D, hexahedra in 3D)
which is the focus of this paper, $N_p = (p+1)^d$, and the basis spans the space
of all polynomials of degree at most $p$ in each variable, denoted by $Q_p$.
The position of an
element $E$ in the mesh $\mathcal{M}$ is fully described by a matrix
$\mathbf{x}_E$ of size $d \times N_p$ whose columns represent the coordinates
of the element {\em control points} (also known as {\em nodes} or element {\em degrees of freedom}).
Given $\mathbf{x}_E$, we introduce the map $\Phi_E:\bar{E} \to \mathbb{R}^d$
whose image defines the physical element $E$:
\begin{equation}
\label{eq_x}
x(\bar{x}) =
   \Phi_E(\bar{x}) \equiv
   \sum_{i=1}^{N_p} \mathbf{x}_{E,i} \bw_i(\bar{x}),
   \qquad \bar{x} \in \bar{E}, ~~ x=x(\bar{x}) \in E,
\end{equation}
where $\mathbf{x}_{E,i}$ denotes the $i$-th column of $\bx_E$, i.e.,
the $i$-th node of element $E$.
To represent the whole mesh, the coordinates of the control points of every
element are arranged in a global vector $\bx$ of size $d \times N_x$ that
stores the coordinates of all node positions and ensures continuity at the
faces of all elements. Here $N_x$ is the global number of control points for
a principal direction.

\subsection{TMOP for \ra}

In this subsection we summarize the main components of the TMOP approach; all
details of the specific method we use are provided in \cite{TMOP2020}.
The input of TMOP is the user-specified transformation matrix $W$, from
reference-space $\bar{E}$ to target element $E_t$, which represents the
ideal geometric properties desired for every mesh point. Note that after
discretization, there will be multiple input transformation matrices $W$ -- one
for every quadrature point in every mesh element.
The construction of this transformation is guided by the fact that any Jacobian
matrix can be written as a composition of four components:
\begin{equation}
\label{eq_W}
W = \underbrace{\zeta}_{\text{[volume]}} \circ
    \underbrace{R}_{\text{[rotation]}} \circ
    \underbrace{Q}_{\text{[skewness]}} \circ
    \underbrace{D}_{\text{[aspect ratio]}}.
\end{equation}
Further details and a thorough discussion on how TMOP's target construction
methods encode geometric information into the target
matrix $W$ is given by Knupp in \cite{knupp2019target}.
In the context of \ra\, the geometric parameters of \eqref{eq_W} are typically
constructed using a discrete PDE solution available on the initial mesh.
As the nodal coordinates change during the optimization process, these discrete
functions are mapped to the updated mesh so that $W$ can be constructed at
each reference point, see Section 4 in \cite{IMR2018}.

\begin{figure}[tb!]
\centerline{
  \includegraphics[width=0.4\textwidth]{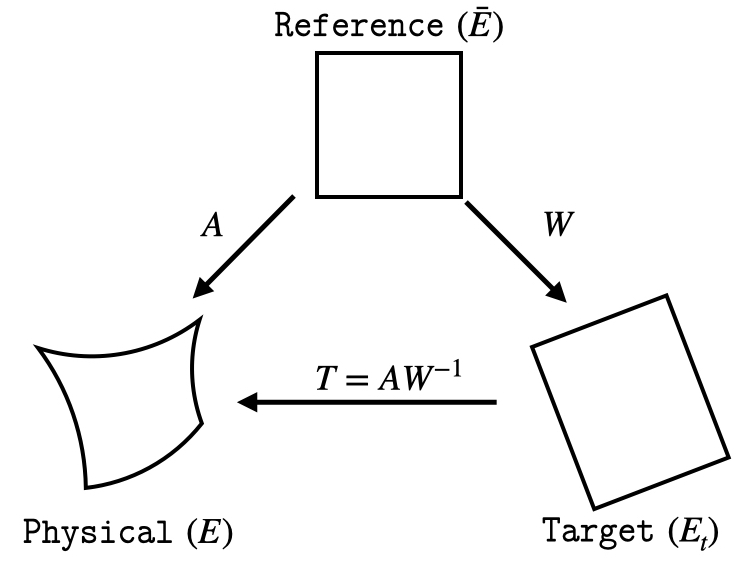}}
\caption{Schematic representation of the major TMOP matrices.}
\label{fig_tmop}
\end{figure}

Using \eqref{eq_x}, the Jacobian $A$ of the mapping $\Phi_E$ from the reference
coordinates $\bar{x} \in \bar{E}$ to the current
physical-space coordinates $x \in E$ is defined as
\begin{equation}
\label{eq_A}
A(\bar{x}) = \frac{\partial \Phi_E}{\partial \bar{x}} =
  \sum_{i=1}^{N_p} \mathbf{\bx}_{E,i} [ \nabla \bw_i(\bar{x}) ]^T,
  \qquad \bar{x} \in \bar{E}.
\end{equation}
Combining \eqref{eq_A} and \eqref{eq_W}, the transformation from the
target coordinates to the current physical coordinates,
see Figure \ref{fig_tmop}, is defined as
\begin{equation}
\label{eq_T}
T = AW^{-1}.
\end{equation}
The matrix $T$ is then used to define a quality metric $\mu(T)$ that measures the
difference between $A$ and $W$ in terms of the geometric parameters of interest
specified in \eqref{eq_W}. For example, $\mu_2=|T|^2/2\tau-1$ is a
\emph{shape} metric that depends on the skew and aspect ratio components, but
is invariant to orientation and volume.  Here, $| T |$ is the Frobenius
norm of $T$ and $\tau=\text{det}(T)$.  Similarly, $\mu_{55}=(\tau-1)^2$ is a
\emph{size} metric that depends only on the volume of the element. There are also
\emph{shape}$+$\emph{size} metrics such as $\mu_7 = | T-T^{-t}|^2$ or
$\mu_9=\tau| T-T^{-t}|^2$ that depend on volume, skew and aspect ratio,
but are invariant to orientation. A list of metrics along with their
theoretical properties can be found in \cite{Knupp2020}.

The quality metric $\mu(T)$ is used to define the TMOP objective function:
for any given vector $\bx$ of node positions, we define
\begin{equation}
\label{eq_F_full}
F(\bx) =
  \sum_{E \in \mathcal{M}} \int_{E_t} \cmb(x_t) \mu(T(x_t)) d x_t +
  \sum_{E \in \mathcal{M}} \int_{E_t} \xi(x - x_0, \delta(x_0)) d x_t,
\end{equation}
where $x=x(\bar{x})=x(\bar{x}(x_t))$ is the element $E$ mapping defined by
$\bx$, $E_t$ is the target element corresponding to the physical element $E$,
$\cmb$ is a user-prescribed spatial weight, and the integral involving
$\xi(x-x_0, \delta(x_0))$ limits the node displacements in a user-specified
manner (a typical choice is $\xi(y,\delta)=|y|^2/\delta^2$) with
$x_0=x_0(\bar{x})=x_0(\bar{x}(x_t))$ being a given initial/reference mesh
position.
The integrals are computed with a standard quadrature rule.
The mesh is optimized by minimizing the objective $F$ through node movement.
More specifically, we solve $\partial F(\bx) / \partial \bx = 0$ with a
nonlinear solver, e.g., Newton's method.
The gradient $\partial F$ and Hessian $\partial^2 F$ of $F$ are discussed in
further detail in Sections \ref{sec_pa_objective_grad} and
\ref{sec_pa_hessian_linsolver}.

\begin{figure}[h!]
\begin{center}
$\begin{array}{ccc}
\includegraphics[height=0.3\textwidth]{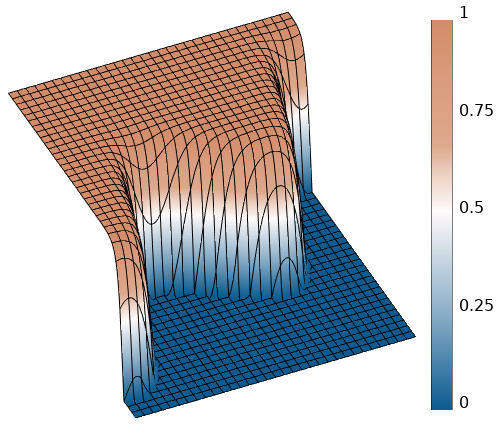} &
\includegraphics[height=0.3\textwidth]{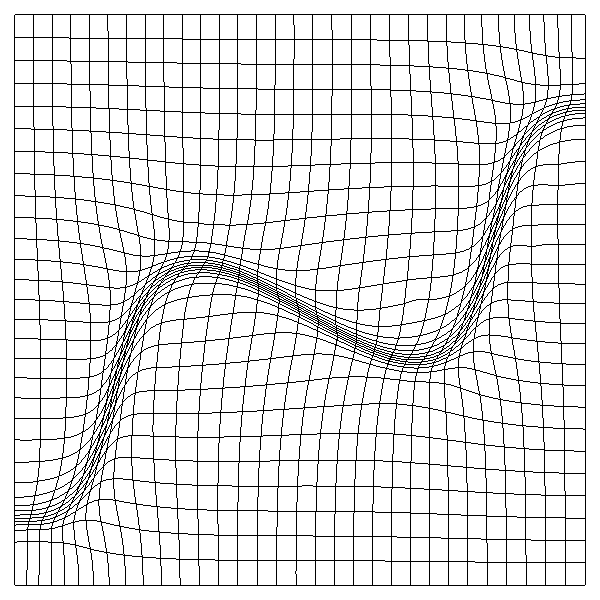} \\
\end{array}$
\end{center}
\vspace{-7mm}
\caption{(left) Material indicator function on the initial uniform mesh
         and (right) the optimized mesh.}
\label{fig_2matind}
\end{figure}

Figure \ref{fig_2matind} shows an example of \ra\ to a
discrete material indicator using TMOP to control the aspect-ratio and the size
of the elements in a mesh. In this example, the material indicator function is
used to define discrete functions for aspect-ratio and size targets in
\eqref{eq_W}, and the mesh is optimized using a \shsz\ metric.

\subsection{Nonlinear Solver}
\label{sec_linsolver}

The optimal nodal locations are determined by minimizing the TMOP objective
function \eqref{eq_F_full} using the Newton's method.
At each Newton iteration, we update the nodal positions as
\begin{eqnarray}
\label{eq_nodal_solve}
\bx^{n+1} &=& \bx^n - \alpha \Delta \bx^n, \\
\label{eq_nodal_dx}
\Delta \bx^n &=& [\partial^2 F(\bx^n)]^{-1} \partial F(\bx^n).
\end{eqnarray}
Here, $\bx^n$ refers to the vector of nodal positions at the $n$-th iteration
during \ra, $\alpha$ is a scalar determined by a
backtracking line-search procedure, and
$\partial^2 F(\bx^n)$ and $\partial F(\bx^n)$ are the Hessian and the gradient,
respectively, associated with the TMOP objective function.
The line-search procedure starts with $\alpha = 1$ and
scales its value by $\frac{1}{2}$ until
$F(\bx^{n+1}) < 1.2 F(\bx^n)$,
$|\partial F(\bx^{n+1})| < 1.2 |\partial F(\bx^n)|$, and
that the determinant of the Jacobian of the transformation at each
quadrature point in the mesh is positive, i.e., $\det(A(\bx^{n+1})) > 0$.
These line-search constraints have been tuned using many numerical experiments
and have demonstrated to be effective in improving mesh quality.
For Newton's method, the Hessian $\partial^2 F(\bx^n)$ is inverted
using the Minimum Residual (MINRES) method with Jacobi preconditioning;
more sophisticated preconditioning techniques can be found in \cite{Roca2022}.
Additionally, the optimization solver iterations \eqref{eq_nodal_solve} are
repeated until the relative $l_2$ norm of the gradient of the objective function
with respect to the current and original mesh nodal positions is
below a certain tolerance $\varepsilon$, i.e.,
$|\partial F(\bx)|/|\partial F(\bx_0)| \leq \varepsilon$.
We set $\varepsilon = 10^{-10}$ for the results presented in the current work.


\section{Partial Assembly for Adaptive Mesh Optimization}
\label{sec_pa}

Traditionally, linear FEM operators are assembled and stored in the form of
sparse matrices, and their action is computed using matrix-vector products.
This approach can become prohibitively expensive when the polynomial degree
$p$ of the basis functions is high, as operator storage, assembly, and
evaluation, per element, scale as $\calO(p^{2d})$, $\calO(p^{3d})$, and
$\calO(p^{2d})$, respectively \cite{MFEM2020}.
In contrast to this full-assembly approach, the high-order finite element community
has demonstrated that
using a tensor-product structure for the basis functions and integration
rule with a matrix-free approach \cite{dfm02, MFEM2020,orszag1979spectral}
reduces the computational complexity of FEM operators such that the
storage, assembly, and evaluation, per element, scale as $\calO(p^{d})$,
$\calO(p^{d})$, and $\calO(p^{d+1})$, respectively.
In this section, we present our approach to extend the notion of partial
assembly for FEM operators to mesh optimization.

\begin{figure}[tb!]
\label{fig_pgb}
\centerline{\includegraphics[width=1\textwidth]{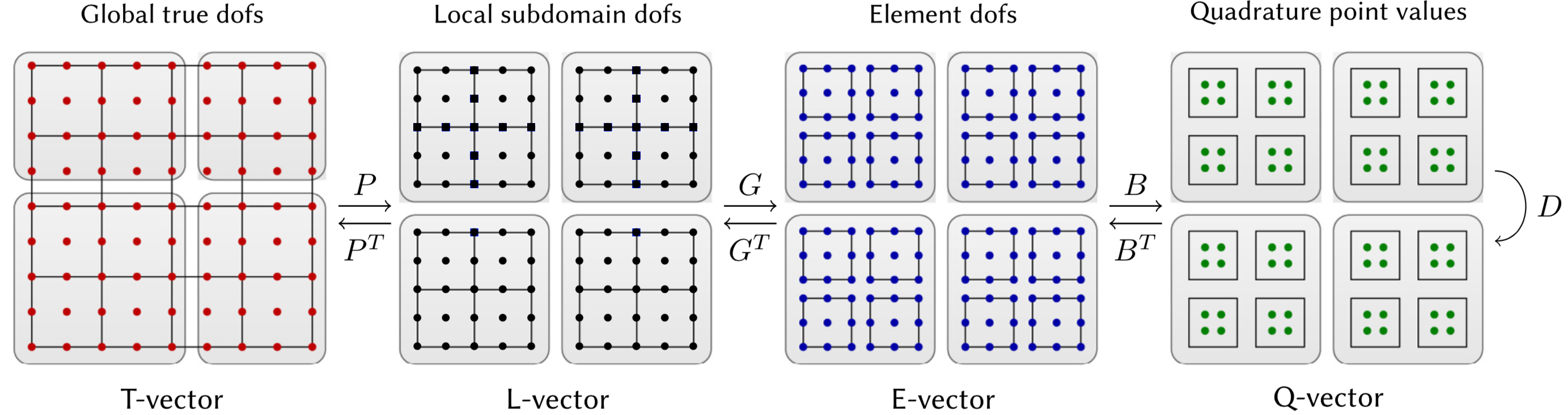}}
\caption{Fundamental finite element operator decomposition, see \cite{MFEM2020,CEED2021}.}
\end{figure}

The PA technique utilizes heavily the fundamental finite element operator
decomposition \cite{CEED2021, MFEM2020}, namely, that any parallel FE operator $A$ can be decomposed as:
\begin{equation}
\label{eq_decompose}
A = P^T G^T B^T D B G P,
\end{equation}
see Figure \ref{fig_pgb}, where
$P$ represents the subdomain restriction operator that maps the global MPI
vector (T-vector) to its local processor values (L-vector),
$G$ represents the element restriction operator that maps the L-vector to its
element values (E-vector),
$B$ represents the basis evaluator that interpolates an E-vector to the
quadrature points inside each element (Q-vector), and finally
$D$ is the operator that defines the quadrature-level computations.
The idea of PA is to compute and store only the result of $D$, and evaluate
the actions of $P, G$ and $B$ on-the-fly.
Furthermore, PA takes advantage of the tensor product structure of the degrees
of freedom and quadrature points on quadrilateral and hexahedral elements to
perform the action of $B$ without storing it as a matrix, as shown later.

In the following subsections we will often refer to the individual components
of the position $x=x(\bar{x})$, namely, $x = (x_1, \dots, x_d)^T$ where $d$
is the dimension and each component is expanded as
$x_a(\bar{x}) = \sum_{i=1}^{N_x} x_{a,i} \bw_i(\bar{x})$, $a = 1, \dots, d$.
Then the discrete vector that contains all nodal positions is
$\bx = \{ x_{a,i} \}$ for all $a = 1, \dots, d$ and $i = 1, \dots, N_x$.
At places we will also redirect the interested reader to the source code of
our implementation. We use the finite element library MFEM \cite{MFEM2020},
specifically version 4.3.


\subsection{Basis Functions and Tensor Products}
\label{sec_pa_basis}

Recall that we represent the mesh positions by \eqref{eq_x}, where on the
reference element we use the scalar basis functions $\{\bw_i\}_{i=1}^{N_p}$.
For quadrilateral and hexahedral elements, we utilize basis functions that are
constructed as tensor products of the standard 1D Lagrange polynomials, e.g.,
for $d=3$ we can write:
\begin{eqnarray}
\label{eqn_basis_tensor}
\bw_i(\bar{x}) =
  \bw_{i_1, i_2, i_3}(\bar{x}_1, \bar{x}_2, \bar{x}_3) =
  \ell_{i_1}(\bar{x}_1) \ell_{i_2}(\bar{x}_2) \ell_{i_3}(\bar{x}_3),
\end{eqnarray}
where $\bar{x} = (\bar{x}_1, \bar{x}_2, \bar{x}_3)$ is a position in reference
space and $\ell_j$ denotes a 1D Lagrange polynomial (or basis function)
in $[0,1]$ of degree $p$.
There are $p+1$ such basis functions, and we will often refer
to this number as $n_i = p+1$.
For simplicity of notation, we will identify the multi-index
$i=(i_1, i_2, i_3)$ with a single index $i$ given by an explicit
formula depending on the ordering. For example, for 3D lexicographic ordering,
\[
  i = 1 + \Big ( i_1 - 1 \Big ) +
          \Big ( i_2 - 1 \Big ) \Big ( p + 1 \Big ) +
          \Big ( i_3 - 1 \Big ) \Big ( p + 1 \Big )^{2}, \quad
  i_1, i_2, i_3 \in \left \{ 1, \dots, p + 1 \right \}.
\]
With the tensor-product decomposition \eqref{eqn_basis_tensor},
the spatial derivatives of the reference basis functions can be written as
\begin{eqnarray*}
\frac{\partial \bw_i(\bar{x})}{\partial \bar{x}_1} =
  \ell'_{i_1}(\bar{x}_1) \ell_{i_2}(\bar{x}_2) \ell_{i_3}(\bar{x}_3),
\frac{\partial \bw_i(\bar{x})}{\partial \bar{x}_2} =
  \ell_{i_1}(\bar{x}_1) \ell'_{i_2}(\bar{x}_2) \ell_{i_3}(\bar{x}_3),
\frac{\partial \bw_i(\bar{x})}{\partial \bar{x}_3} =
  \ell_{i_1}(\bar{x}_1) \ell_{i_2}(\bar{x}_2) \ell'_{i_3}(\bar{x}_3).
\end{eqnarray*}

In addition, we always use quadrature rules that are constructed from
products of 1D quadrature rules.
More specifically, suppose we use a quadrature rule with $n_q$ quadrature points
in each direction, having a total of $N_q = n_q^d$ points in each element.
Then a 3D quadrature point $\bchi_q$
for $q = (q_1, q_2, q_3)$ can be represented as
\[
  \bchi_q =
  \bchi_{q_1, q_2, q_3} = (\chi_{q_1}, \chi_{q_2}, \chi_{q_3}),
\]
where $\{ \chi_{j} \}$ for
$j = 1, \ldots, n_q$ are the points of the 1D (e.g. Gaussian) rule.
When both the basis functions and the quadrature rule have tensor product
structure, we can use tensor contractions to transfer data between degrees of
freedom and quadrature points, leading to efficient finite element calculations
as described in the following sections.

Once a quadrature rule is selected, we define 1D reference matrices of
size $n_q \times n_i$:
\begin{equation}
\label{eq_B_G_1D}
B^{\oneD}_{qi} = \ell_i(\chi_q), \quad G^{\oneD}_{qi} = \ell_i'(\chi_q), \quad
i = 1, \dots, n_i, \quad q = 1, \dots, n_q,
\end{equation}
where $\ell_i$ is a 1D basis function and $\chi_q$ is a quadrature point of the
1D quadrature rule.


\subsection{Quadrature Level Calculations}
\label{sec_pa_quadp}

The primary calculation at a quadrature point is that of
the Jacobian matrix $A$, see \eqref{eq_A}.
For a reference point $\bar{x}$ the entries of $A=A(\bar{x})$ can be
written as:
\[
A_{ab} = \frac{\partial x_a}{\partial \bar{x}_b} =
         \sum_{i=1}^{N_p}
         x_{a,i} \frac{\partial \bw_i(\bar{x})}{\partial \bar{x}_b}, \quad
         a,b = 1, \dots, d.
\]
Since a given element contains $\calO(p^d)$ degrees of freedom (DOFs) and
$\calO(p^d)$ quadrature
points, computing $A$ at all quadrature points would be of complexity
$\calO(p^{2d})$ assuming we precompute and store the terms
${\partial \bw_i(\boldsymbol{\xi}_q)}/{\partial \bar{x}_b}$.
When we utilize 1D tensor contractions, the entries of $A_{ab}$ at all
quadrature points in the element can be computed by
(example for the 3D case when $b=2$):
\[
A_{a2}(\bchi_q) =
\left( \sum_{i_1} B^{\oneD}_{q_1, i_1}
\left( \sum_{i_2} G^{\oneD}_{q_2, i_2}
\left( \sum_{i_3} B^{\oneD}_{q_3, i_3} x_{a,i_1,i_2,i_3}
   \right) \right) \right) =
   \left[
   B^{\oneD} \otimes G^{\oneD} \otimes B^{\oneD} \, \bx_{E,a}
   \right]_{q_1,q_2,q_3}
\]
where $\bx_{E,a}=\{x_{a,i}\}_{i=1}^{N_p}$.
Here the $\otimes$ symbol denotes the tensor outer product which means that the
matrices will be applied as a sequence of multiplications as given by the above
formula.
The important point here is that each contraction is of complexity
$\calO(p^{d+1})$
(as $\bx_{E,a}$ is a tensor of rank $d$), resulting in an overall complexity of
$\calO(p^{d+1})$.
This is a significant improvement over $\calO(p^{2d})$, especially for high $p$.
One complication with tensor contractions is that the implementation has
to take care of arranging the vectors $\bx_E$ in a suitable tensor form.

The rest of the quadrature level calculations are independent of the finite
element discretization, i.e., they do not involve the finite element basis
functions $\{\bw_i\}$. Here the notion of sum factorization does not apply;
computational gains can be achieved on GPU devices by proper batching
over many quadrature points.
The major quadrature level calculations are the following:
\begin{itemize}
\item Given physical coordinates of a point $x$,
      construct the target matrix $W(x)$ and compute $\det(W(x))$.
      In the case of adaptivity, additional simulation-specific values also
      provided as input to the computation of $W$.
\item Given a $d \times d$ matrix $T$,
      compute the mesh quality metric the $\mu(T)$,
      its first derivative $\partial \mu / \partial T$ which is a $d \times d$
      matrix, and its second derivative $\partial^2 \mu / \partial T^2$ which
      is a $d \times d \times d \times d$ tensor.
      These are derived using standard matrix algebra,
      e.g. \cite{petersen2012matrix}.
\end{itemize}


\subsection{Objective Function and First Derivative}
\label{sec_pa_objective_grad}

In this section we consider the TMOP objective function of the form
\eqref{eq_F_full} and its first derivative. In this section we focus on the
first term in \eqref{eq_F_full} while the second term will be discussed later
in Section \ref{sec_pa_limiting}.

To evaluate the objective function $F(\bx)$ one can use the following
decomposition similar to \eqref{eq_decompose}
\[
F(\bx) = {\bf 1}^T D_0 ( B G P \bx ),
\]
where $\bf{1}$ is a vector of ones and $D_0$ evaluates, at all quadrature
points, the integrand of the first term in \eqref{eq_F_full} after changing the
variables in the integral from $E_t$ to $\bar{E}$ and applying the quadrature
rule:
\begin{equation}
\label{D0_expression}
D_0(A) = w_q \det(W) \cmb(x_t) \mu(A W^{-1}).
\end{equation}
Here $w_q$ is the quadrature point weight,
while $\cmb(x_t), \det(W)$, and $W$ are the
precomputed data for the current quadrature point,
and $A$ is the Jacobian
matrix at the current quadrature point computed by evaluating $B G P \bx$.

The main speedup in the calculation of the objective function comes from proper
batching of the computation of the Jacobian $A$ over all quadrature points, as
explained in Section \ref{sec_pa_quadp}. Once $T=A W^{-1}$ and $\mu(T)$ are
computed at all quadrature points, the final integral is readily available.

The computation of the gradient $\partial F / \partial \bx$ is improved by
utilizing the PA technique. For simplicity, in the formulas below we
assume that $W=I$ (the target elements coincide with the reference one),
although this is generally not the case as explained in
Section \ref{sec_pa_remap}.
Then the first derivative of the objective $F$ with respect of the node
$x_{a,i}$ for each element is:
\begin{equation}
\label{eq_vm_pa_grad}
\begin{split}
\frac{\partial F(\bx)}{\partial x_{a,i}}
  & = \int_{\bar{E}}
      \frac{\partial \mu}{\partial T(\bx)} :
      \frac{\partial T(\bx)}{\partial x_{a,i}} d\bar{x}
    =  \int_{\bar{E}} \frac{\partial \mu}{\partial T} :
    \left( \frac{\partial A}{\partial x_{a,i}} \right) d\bar{x} \\
  & =
  \sum_{q=1}^{N_q} z_q
    \bigg[ \sum_{m,n = 1}^d \frac{\partial \mu}{\partial T_{mn}}
           \delta_{m, a} \frac{\partial \bw_i(\bchi_q)}{\partial \bar{x}_n}
    \bigg], \quad a = 1, \dots, d, ~~ i = 1, \dots, N_x,
\end{split}
\end{equation}
where $z_q$ is the quadrature point weight and $\delta$ is the standard
Kronecker delta. The dependence of $x$ in
\eqref{eq_vm_pa_grad} comes through the nonlinearity of $\mu(T)$.
With PA we efficiently compute all quadrature points together, as a result of a
tensor contraction. We first form a $D$ matrix of size
$n_q \times n_q \times n_q$ in each element that stores all quadrature data,
e.g., the derivatives $\partial \mu / \partial T$.
This quadrature-based calculation is of complexity $\mathcal{O}(p^{d})$.
Then $D, \bx, B, G$ are contracted appropriately to obtain the final derivative
vector, with complexity $\mathcal{O}(p^{d+1})$.
These contraction calculations are complicated and can be found
in the mfem-4.3 source in files \code{fem/tmop/tmop{\_}pa{\_}p2.cpp} and
\code{fem/tmop/tmop{\_}pa{\_}p3.cpp} for 2D and 3D, respectively.
Once the gradient of the TMOP objective function is computed on an
element-by-element basis (E-vector), it is mapped to a global T-vector, see
Figure \ref{fig_pgb}, to ensure that the gradients are consistent for
DOFs that are shared between neighboring elements.


\subsection{Second Derivative and Linear Solver}
\label{sec_pa_hessian_linsolver}

Again assuming $W = I$, the second derivative of $F$ for the
nodes $x_{a,i}$ and $x_{b,j}$ is:
\begin{equation}
\label{eq_vm_pa_hess}
\begin{split}
\frac{\partial^2 F(\bx)}{\partial x_{b,j} \partial x_{a,i}}
  = &
  \int_{\bar{E}} \frac{\partial}{\partial x_{b,j}} \bigg[
    \frac{\partial \mu}{\partial T} :
    \left( \frac{\partial A}{\partial x_{a,i}} \right) \bigg] d\bar{x} \\
  = &
  \sum_{q=1}^{N_q} z_q \bigg[
    \sum_{m,n = 1}^d \sum_{o,p = 1}^d
      \frac{\partial^2 \mu}{\partial T_{op} \partial T_{mn}}
         \delta_{o, b} \frac{\partial \bw_j(\bchi_q)}{\partial \bar{x}_p}
         \delta_{m, a} \frac{\partial \bw_i(\bchi_q)}{\partial \bar{x}_n}
  \bigg], \\
  & \quad a,b = 1, \dots, d, ~~ \quad i,j = 1, \dots, N_x,
\end{split}
\end{equation}
where $z_q$ is the quadrature point weight and $\delta$ is the
standard Kronecker delta.
Assembling and applying the Hessian matrix $\partial^2 F$ are the two most
expensive computations in our algorithm.
The classical (full assembly) approach forms a sparse matrix over all DOFs in
the problem, requiring to store $\mathcal{O}(p^{2d})$ entries per element.
Computing these entries with standard nested loops requires
$\mathcal{O}(p^{3d})$ operations per element, while applying the sparse matrix
to a vector is complexity $\mathcal{O}(p^{2d})$.
Since $\partial^2 F$ depends on the current mesh positions (due to the
nonlinearity of $\mu$ with respect to $T$), the assembly must
be performed at every Newton iteration. For higher mesh orders $p$ these
computations become prohibitively expensive, especially in 3D.

The use of PA for $\partial^2 F$ is critical, as it avoids the formation of a
global sparse matrix.
In this case the assembly is replaced by pre-computing and storing the
quadrature-specific data at all quadrature points, with complexity and
storage of $\mathcal{O}(p^{d})$ per element.
The action of $\partial^2 F$ is performed element-by-element, using tensor
contractions involving the quadrature data, the DOF vector $\bx$, and the 1D
matrices $B$ and $G$.
This PA-based action is of complexity $\mathcal{O}(p^{d+1})$ per element.
The complete calculation can be seen in mfem-4.3's source files
\code{fem/tmop/tmop{\_}pa{\_}h2m.cpp} and \code{fem/tmop/tmop{\_}pa{\_}h3m.cpp}
for 2D and 3D, respectively.
The assembly of all quadrature point data is in the files
\code{fem/tmop/tmop{\_}pa{\_}h2s.cpp} and \code{fem/tmop/tmop{\_}pa{\_}h3s.cpp}
for 2D and 3D, respectively.
The inversion of $\partial^2 F$ is performed by an iterative solver that uses
the PA-based action of the operator.
Our default choice is the Minimum Residual (MINRES) method,
as $\partial^2 F$ is symmetric but not necessarily positive-definite.
Preconditioning for matrix-free inversion is a substantial challenge and an
active area of research \cite{franco2020high}.
We have the option to perform Jacobi preconditioning, as the diagonal of
$\partial^2 F$ can be computed through tensor contractions without having the
global matrix; these algorithms can be found in files
\code{fem/tmop/tmop{\_}pa{\_}h2d.cpp} and \code{fem/tmop/tmop{\_}pa{\_}h3d.cpp}
for 2D and 3D, respectively.


\subsection{Limiting of Node Displacements}
\label{sec_pa_limiting}

The term involving the limiting function $\xi(x-x_0, \delta(x_0))$ in
\eqref{eq_F_full} is used to restrict the node displacements in
certain regions of the domain during mesh optimization,
see Section 3.2 of \cite{TMOP2020}.
A simple example of a node limiting function is
$\xi(x-x_0, \delta(x_0)) = (x-x_0)^2 / \delta^2$, which
would start to dominate the objective function once the displacements go
above the local $\delta$ value.
Computing this term and its derivatives through PA does not present any
difficulty.  For example, the second derivative of this term is:
\begin{equation}
\label{eq_pa_limiting}
\begin{split}
  \frac{\partial^2}{\partial x_{b,j} \partial x_{a,i}}
    \int_{E_t} \xi(x-x_0, \delta(x_0)) d x_t &=
  \int_{\bar{E}} \det(W) \frac{\partial^2 \xi}{\partial x_b \partial x_a}
    \frac{\partial x_b}{\partial x_{b,j}}
    \frac{\partial x_a}{\partial x_{a,i}} d\bar{x} \\
  & = \sum_{q=1}^{N_q} z_q \det(W(\bar{x}_q))
    \frac{\partial^2 \xi}{\partial x_b \partial x_a}
    \bw_j(\bar{x}_q) \bw_i(\bar{x}_q),
\end{split}
\end{equation}
which resembles a standard FE mass operator.
Dependence on the current mesh positions $x$ in \eqref{eq_pa_limiting} appears
when $\xi$ is chosen as a nonlinear function with respect to $x-x_0$.
The PA-based action is formed by contracting $B$ and $\bx$ with the matrix $D$
that contains all the quadrature data needed in \eqref{eq_pa_limiting}.
For example, in 3D, the action of this operator to a vector $\bx$
(rearranged as a tensor $X$) is computed as
\[
B^{\oneD^T} \otimes B^{\oneD^T} \otimes B^{\oneD^T} \otimes
D \otimes B^{\oneD} \otimes B^{\oneD} \otimes B^{\oneD} X.
\]
The implementation of the above computation can be found in the following
mfem-4.3 source files: \code{fem/tmop/tmop{\_}pa{\_}h2m{\_}c0.cpp} and
\code{fem/tmop/tmop{\_}pa{\_}h3m{\_}c0.cpp} for 2D and 3D, respectively.


\subsection{Adaptivity and Field Transfer}
\label{sec_pa_remap}

When TMOP is used to achieve r-adaptivity, the definitions of the target
matrices contain information about discrete simulation fields \cite{TMOP2020}.
For illustration purposes, we denote such simulation fields by $\eta(x)$, which
for our purposes is always a FE function.
The evaluation of $W$ requires the evaluation of $\eta(x)$, which is optimized
by tensor contractions, that is, the $\eta$ values at all quadrature points are
obtained through $d$ contractions of the $B^{\oneD}$ matrices.

The target matrices $W$ becomes space-dependent through its $\eta$ dependence.
Since $\partial W / \partial x \neq 0$, the derivatives in equations
\eqref{eq_vm_pa_grad}, \eqref{eq_vm_pa_hess} and \eqref{eq_pa_limiting} become
more complicated, as well as their PA implementation.
In our experience, adding the extra derivative
terms does not change the results in a significant way,
and we don't include these terms in our current implementation.
The dependence of $W$ on $x$ is still accounted by the algorithm, as the
targets are recomputed after every position change in the nonlinear iteration.

Since finite element fields are only defined with respect to the
initial mesh $\mathcal{M}_0$, remap procedures are needed to obtain their
values on the different meshes obtained during the mesh optimization iterations \cite{IMR2018,mittal2019nonconforming}.
Our GPU implementation utilizes the so-called advection approach, where remap is
achieved by solving an advection PDE in pseudo-time.
This approach is entirely based on standard finite element operations,
enabling the use of sum factorization and optimized matrix-free kernels.
By defining a mesh velocity $v = x - x_0$,
the remap of a function $\eta(x)$ is posed as the following PDE in
pseudo-time $\tau \in [0,1]$:
\[
  \frac{d \eta(x_{\tau}, \tau)}{d \tau} =
    v \cdot \nabla \eta(x_{\tau}, \tau), \quad
  x_{\tau} = x_0 + \tau v, \quad
  \eta(x_0, 0) = \eta_0(x_0).
\]
More details about the mathematical formulation can be found in Section 4.2 of
\cite{IMR2018}.
Solving the above PDE in a matrix-free manner requires to define the actions of
a mass operator and an advection operator.
The mass operator is applied many times per time step, as it is used by a
conjugate gradient linear solver.
The advection operator is applied once per time step, to form the right-hand-side
of the linear system.
The actions of these operators are standard FE kernels, and we directly use
their optimized PA implementations from MFEM.


\section{Numerical Results}
\label{sec_results}

In this section we present several numerical examples and benchmarks
demonstrating the improvement in computational efficiency due to use of partial
assembly on GPU-based architectures in comparison to CPU computations.


\subsection{Kershaw Benchmark}
\label{subsec_kershaw}

Below we propose a performance benchmark for optimization of
high-order meshes. The setup is based on the Kershaw meshes
introduced in \cite{kershaw1981differencing}.
The test is designed so that both the initial (deformed) mesh and the final
(optimized) mesh are straightforward to reproduce.
This allows to compare the performance of different optimization methods on
a problem with well-defined initial configuration and final output.

\begin{figure}[t!]
\begin{center}
$\begin{array}{cccc}
\includegraphics[height=0.24\textwidth]{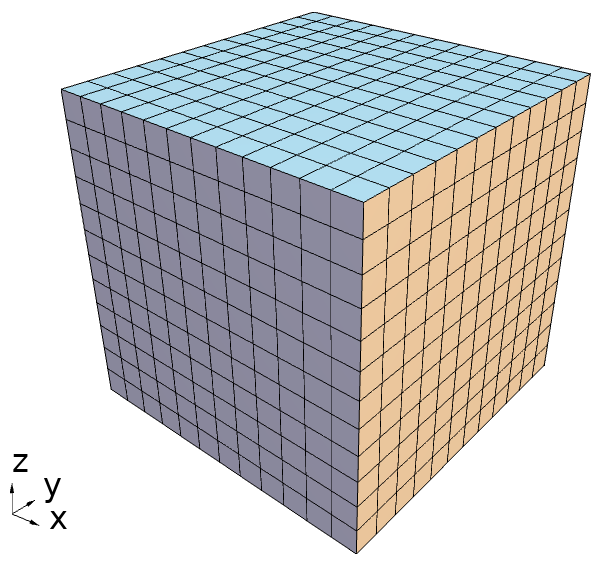} &
\includegraphics[height=0.24\textwidth]{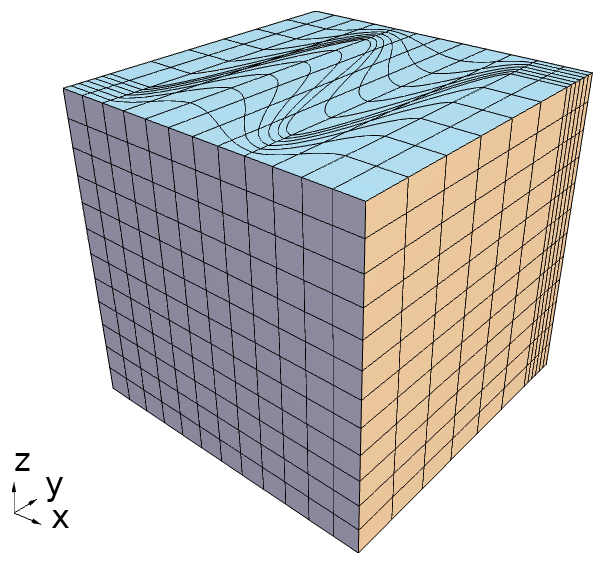} &
\includegraphics[height=0.24\textwidth]{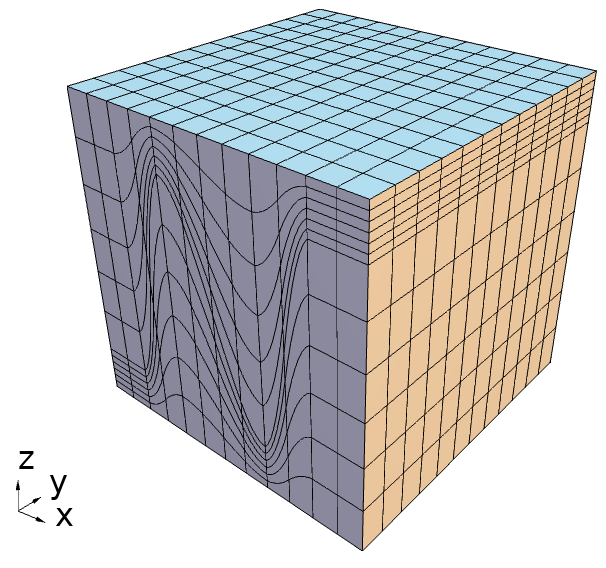} &
\includegraphics[height=0.24\textwidth]{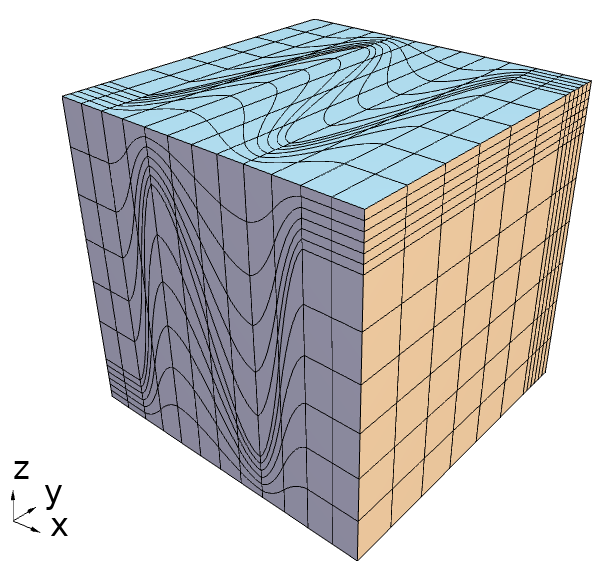} \\
\textrm{(a) $\epsy = \epsz = 1$} &
\textrm{(b) $\epsy = 0.3$, $\epsz = 1$} &
\textrm{(c) $\epsy = 1$, $\epsz = 0.3$} &
\textrm{(d) $\epsy = \epsz = 0.3$}
\end{array}$
\end{center}
\vspace{-7mm}
\caption{Kershaw meshes for various anisotropy parameters.}
\label{fig_kershaw_meshes}
\end{figure}

The initial mesh (that will be optimized) is obtained by an analytic
deformation of a uniform Cartesian mesh. The deformation curves and distorts
the elements, allowing to quantify the performance of the method in terms of
speed and accuracy for meshes of different orders and element counts.
The Kershaw meshes \cite{kershaw1981differencing, CEEDMS36} are parameterized by
two anisotropy parameters $\epsy, \epsz \in (0, 1]$ such that
the meshes are uniform for $\epsy = \epsz = 1$ and become
increasingly anisotropic as $\epsy$ an $\epsz$ decrease,
see Figure \ref{fig_kershaw_meshes}.
To obtain the Kershaw mesh, the $x$-axis of the mesh is decomposed into 6
equally sized layers, and elements in the leftmost and rightmost layers are
modified to have aspect-ratios $1/\epsy$ and $1/\epsz$.
The high aspect-ratio elements are placed in the opposite corners,
while the intermediate layers
are smoothly transitioned with a piecewise quintic function.
The generating C++ code is provided in \ref{sec_appendix} for convenience.
Due to the problem setup, the number of elements in the
$x$-direction must be an integer multiple of 6,
while in the $y$- and $z$-directions must be multiples of 2.

\paragraph{Baseline Wall Time}
First we report baseline time-to-solution for a complete mesh optimization
computation, that is, evolving an initially deformed mesh to the ideal uniform
configuration.
The deformed meshes are obtained by $\epsy = \epsz = 0.3$ and the resolution
is fixed to $N_e = 24 \times 24 \times 24$ number of elements.
We present timings for different mesh orders, different
solver strategies (Newton's method with and without preconditioner),
different architectures (CPU versus GPU), for both full assembly and partial
assembly.
Full assembly computations on GPU are not performed, as storing
big sparse matrices in GPU memory is not feasible.
For all experiments the quadrature order is fixed at 8,
resulting in $N_q = 9^3$ quadrature points per element,
maximum number of linear solver iterations per Newton iterations is 50,
and $\mu_{303} = \frac{|T|^2}{3\tau^(2/3)}-1$ is used for shape optimization.
The results presented here were obtained on Lassen,
a Livermore Computing supercomputer, that has IBM Power9 CPUs
(792 nodes with 44 CPU cores per node) and NVIDIA V100 GPUs (4 per node).
All reported computations utilize a single machine node. The CPU runs use 36
CPU cores, while the GPU runs utilize 4 CPU cores with 1 GPU per core.

\begin{table}[h!]
\begin{center}
$\begin{array}{cc}
\hspace{-5mm}
\begin{tabular}{| c | c | c | c | c |}
\hline
 & \multicolumn{4}{c |}{Unique Degrees of Freedom}\\
 \hline
  & $p = 1$ & $p = 2$ & $p = 3$ & $p = 4$  \\
\hline
 & 46,875 & 352,947 & 1,167,051 & 2,738,019 \\
\hline
\hline
 & \multicolumn{4}{c |}
{Quadrature points per core
         $\left(\frac{N_e N_q}{\text{\# cores}}\right)$}\\
\hline
CPU & \multicolumn{4}{c|}{279,936} \\
\hline
GPU & \multicolumn{4}{c|}{2,519,424} \\
\hline
\end{tabular}
&
\begin{tabular}{| c | c | c | c | c |}
\hline
 & \multicolumn{4}{c |}{Time to solution (sec)}\\
\hline
 & $p = 1$ & $p = 2$ & $p = 3$ & $p = 4$  \\
\hline
CPU\textsuperscript{FA} &  4.9 & 50.9 & 530.9 & 2838.5 \\
\hline
CPU\textsuperscript{FA*} &  2.9 & 31.1 & 489.6 & 2868.8 \\
\hline
CPU\textsuperscript{PA} &  24.1 & 78.3 & 152.3 & 252.9 \\
\hline
CPU\textsuperscript{PA*} & 18.0  & 41.0 & 128.5 & 298.0  \\
\hline
GPU\textsuperscript{PA} & 0.5 & 1.7 & 4.7 & 7.8 \\
\hline
GPU\textsuperscript{PA*} & 0.4  & 0.9 & 3.9  & 8.5 \\
\hline
\hline
 & \multicolumn{4}{c |}
 {Speedup (GPU\textsuperscript{PA*} vs CPU\textsuperscript{PA*})}\\
\hline
 & \textbf{42$\times$} & \textbf{43$\times$} & \textbf{32$\times$} & \textbf{35$\times$}\\
\hline
\end{tabular}
\\
\textrm{(a) } &
\textrm{(b) } \\

\begin{tabular}{| c | c | c | c | c |}
\hline
 & \multicolumn{4}{c |}{Total Newton Iterations}\\
\hline
  & $p = 1$ & $p = 2$ & $p = 3$ & $p = 4$\\
\hline
CPU\textsuperscript{FA} &  19 & 29 & 44 & 76 \\
\hline
CPU\textsuperscript{FA*} &  11 & 18 & 41 & 76 \\
\hline
CPU\textsuperscript{PA} &  19 & 29 & 44 & 70 \\
\hline
CPU\textsuperscript{PA*} & 11 & 18 & 40 & 75  \\
\hline
GPU\textsuperscript{PA} & 19 & 29 & 44 & 67  \\
\hline
GPU\textsuperscript{PA*} & 11 & 18 & 41 & 70 \\
\hline
\end{tabular}
&
\begin{tabular}{| c | c | c | c | c |}
\hline
 & \multicolumn{4}{c |}{Total MINRES Iterations}\\
\hline
  & $p = 1$ & $p = 2$ & $p = 3$ & $p = 4$\\
\hline
CPU\textsuperscript{FA} &  284 & 1108 & 2064 & 3327 \\
\hline
CPU\textsuperscript{FA*} &  203 & 507 & 1548 & 3459 \\
\hline
CPU\textsuperscript{PA} &  284 & 1109 & 2062 & 3209 \\
\hline
CPU\textsuperscript{PA*} & 203 & 507 & 1560 & 3410 \\
\hline
GPU\textsuperscript{PA} & 284 & 1107 & 2060 & 3139\\
\hline
GPU\textsuperscript{PA*} & 203 & 507 & 1536 & 3110\\
\hline
\end{tabular} \\
\textrm{(c) } &
\textrm{(d) } \\
\end{array}$
\end{center}
\vspace{-7mm}
\caption{For meshes of different orders ($p$), we compare
(a) unique degrees of freedom and total quadrature points per core,
(b) total time to solution,
(c) total Newton iterations, and
(d) total MINRES iterations.
The symbol * denotes the use of a preconditioner for the MINRES solve.}
\label{tab_kershaw_baseline}
\end{table}

\begin{figure}[h!]
\begin{center}
$\begin{array}{ccc}
\includegraphics[height=0.28\textwidth]{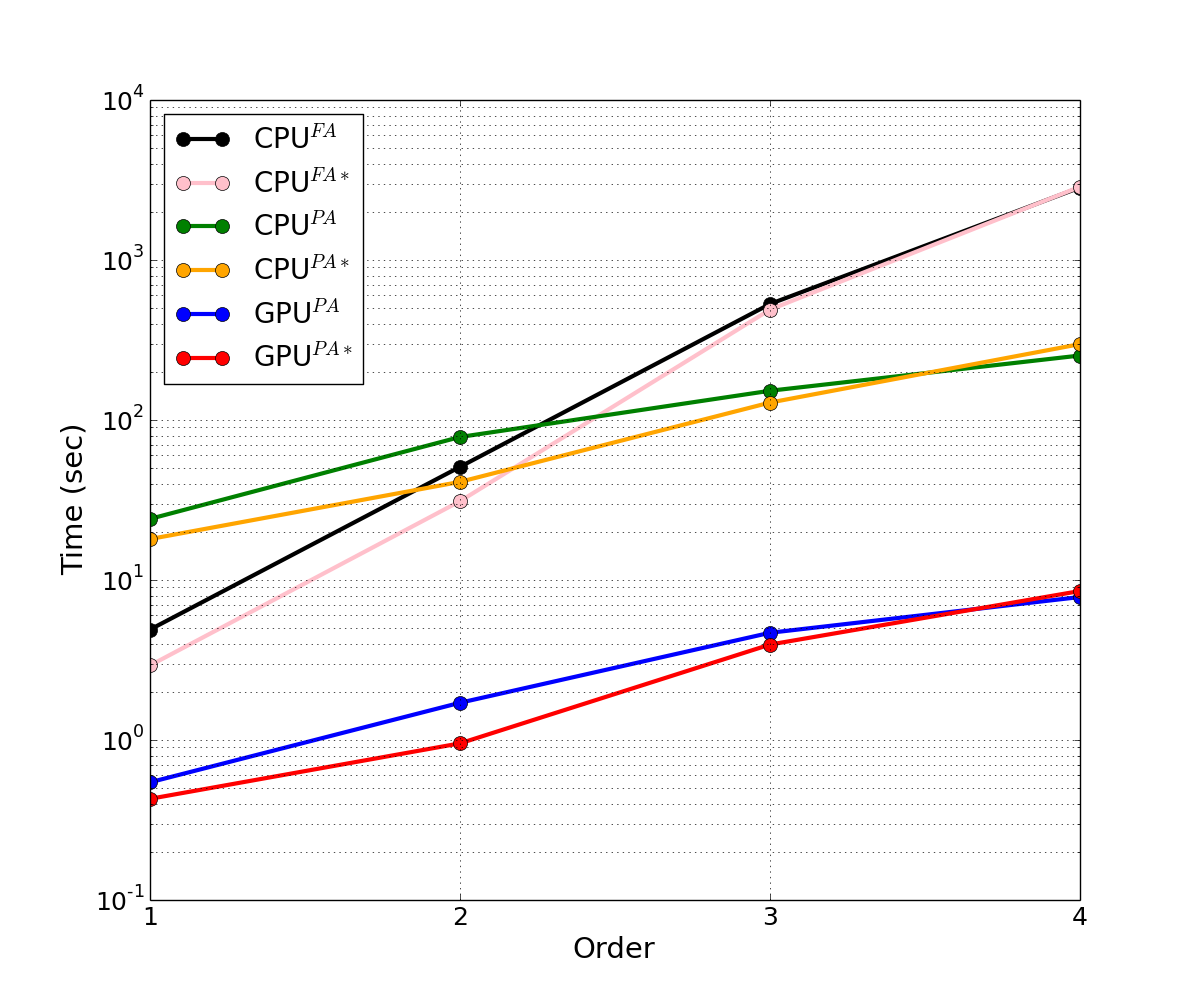} &
\hspace{-10mm}
\includegraphics[height=0.28\textwidth]{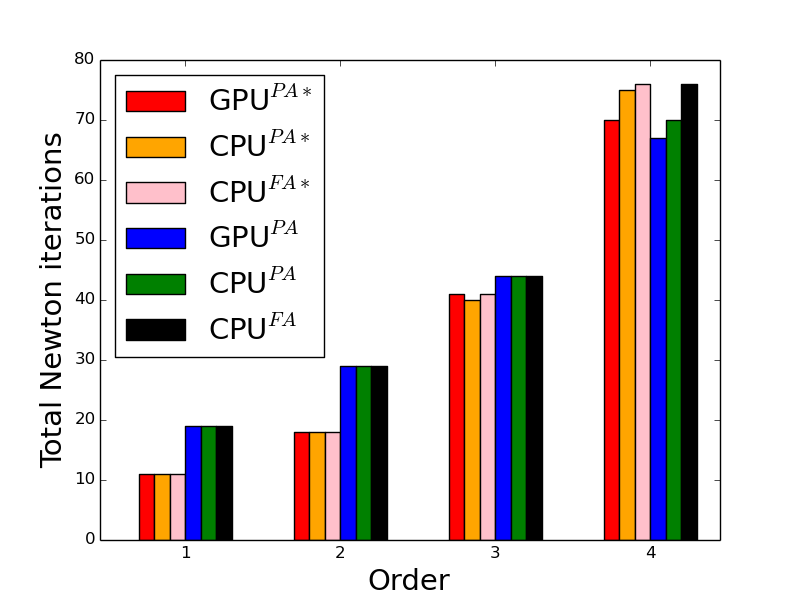} &
\hspace{-10mm}
\includegraphics[height=0.28\textwidth]{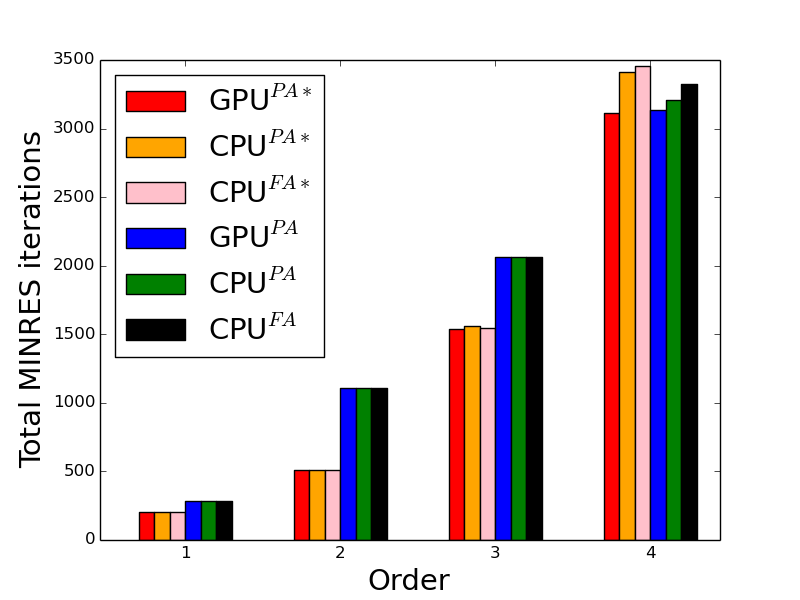} \\
\textrm{(a) Total time to solution} &
\hspace{-10mm}
\textrm{(b) Newton iterations} &
\hspace{-10mm}
\textrm{(c) MINRES iterations}
\end{array}$
\end{center}
\vspace{-7mm}
\caption{Comparison of different solver strategies and architectures.}
\label{fig_kershaw_baseline}
\end{figure}

\begin{figure}[h!]
\begin{center}
$\begin{array}{ccc}
\includegraphics[height=0.2\textwidth]{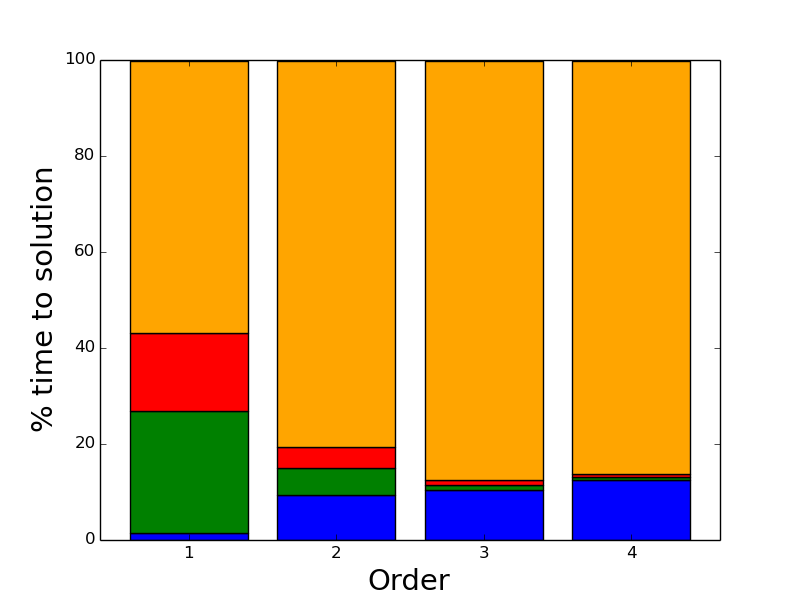} &
\includegraphics[height=0.2\textwidth]{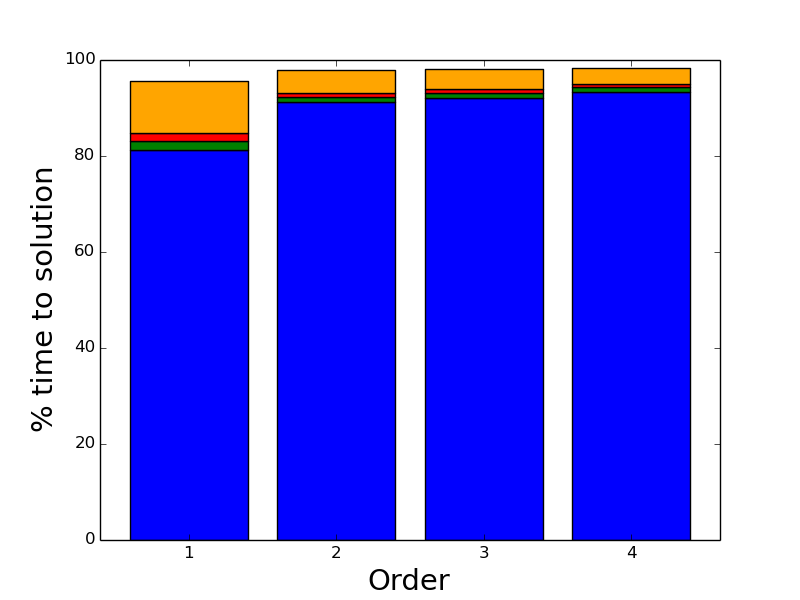} &
\includegraphics[height=0.2\textwidth]{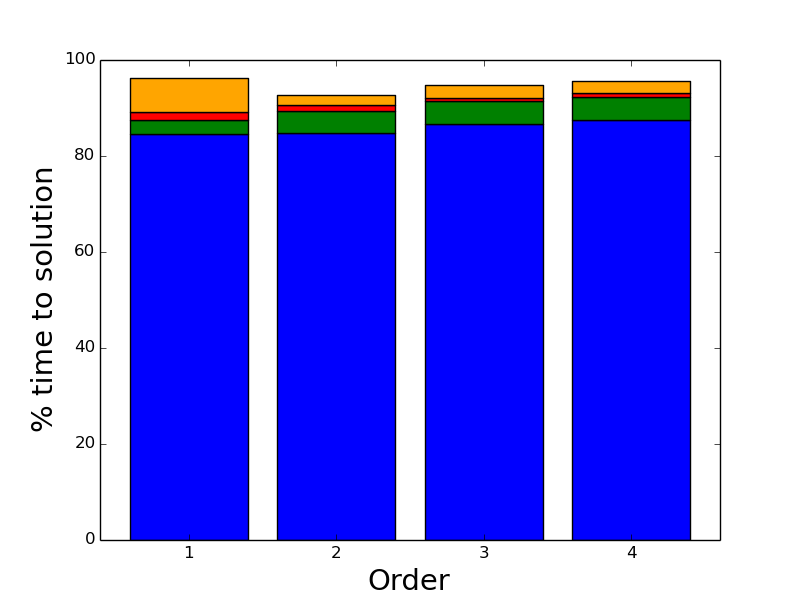} \\
\textrm{(a) CPU\textsuperscript{FA}} &
\textrm{(b) CPU\textsuperscript{PA}} &
\textrm{(c) GPU\textsuperscript{PA}} \\
\includegraphics[height=0.2\textwidth]{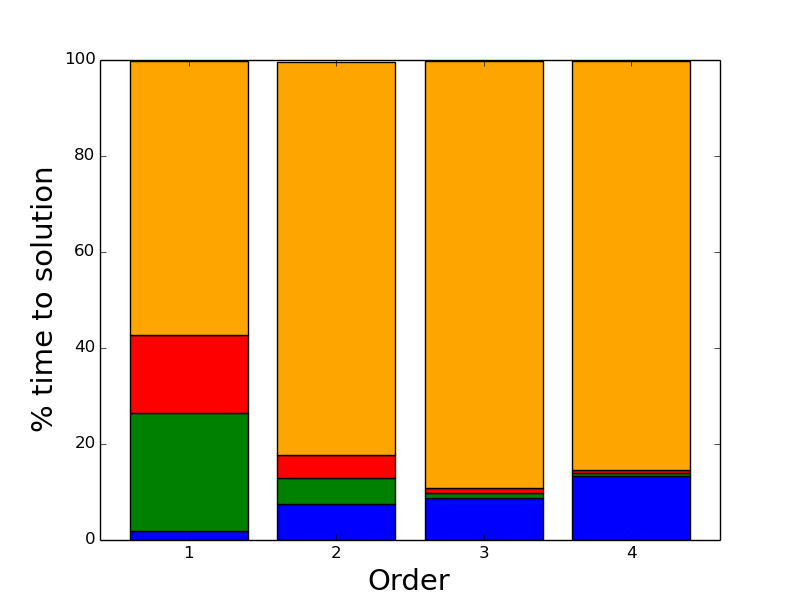} &
\includegraphics[height=0.2\textwidth]{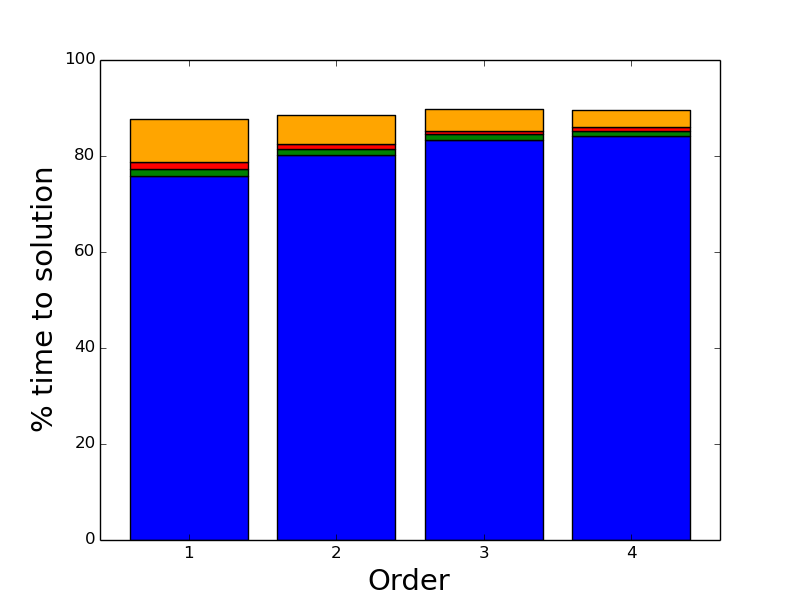} &
\includegraphics[height=0.2\textwidth]{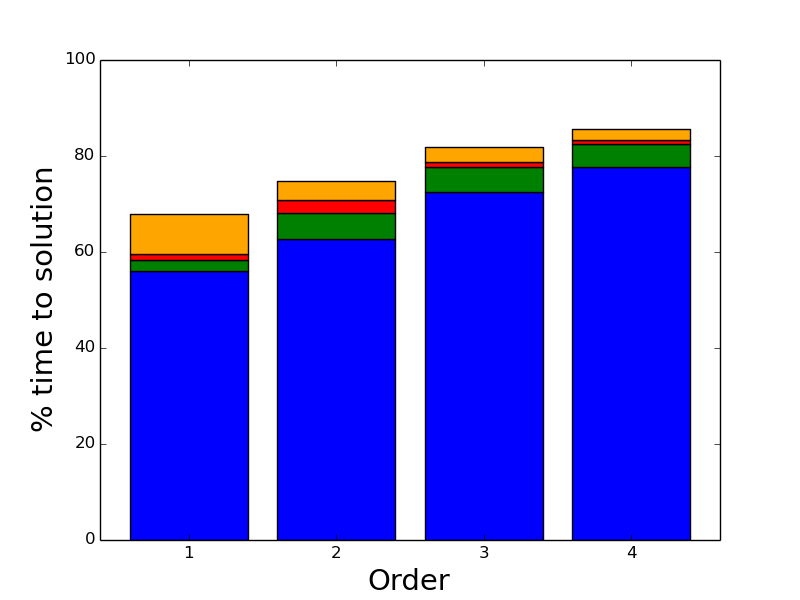} \\
\textrm{(d) CPU\textsuperscript{FA*}} &
\textrm{(e) CPU\textsuperscript{PA*}} &
\textrm{(f) GPU\textsuperscript{PA*}} \\
\multicolumn{3}{c}
{\includegraphics[height=0.05\textwidth]{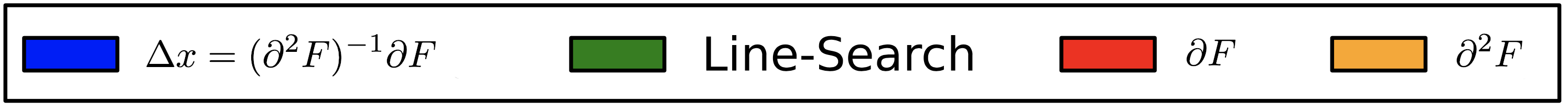}}
\end{array}$
\end{center}
\vspace{-7mm}
\caption{Percentage of the total time spent in different Kernels
for different solver strategies and architectures.}
\label{fig_kershaw_baseline_split}
\end{figure}

Table \ref{tab_kershaw_baseline} compares the unique degrees of freedom
and total quadrature points per core, total time to solution, Newton iterations,
and MINRES iterations, for different mesh orders ($p=1,\ldots, 4$) and
different solver strategies. The corresponding data is also shown in
Figure \ref{fig_kershaw_baseline}. We note that the use of GPUs leads to a
30-40$\times$ speed up in comparison to CPUs. We also quantify the total time spent in
each of the main kernels (gradient, Hessian, etc.) during mesh optimization in
Figure \ref{fig_kershaw_baseline_split}.

\paragraph{Strong scaling}

\begin{figure}[h!]
\begin{center}
\includegraphics[height=0.6\textwidth]{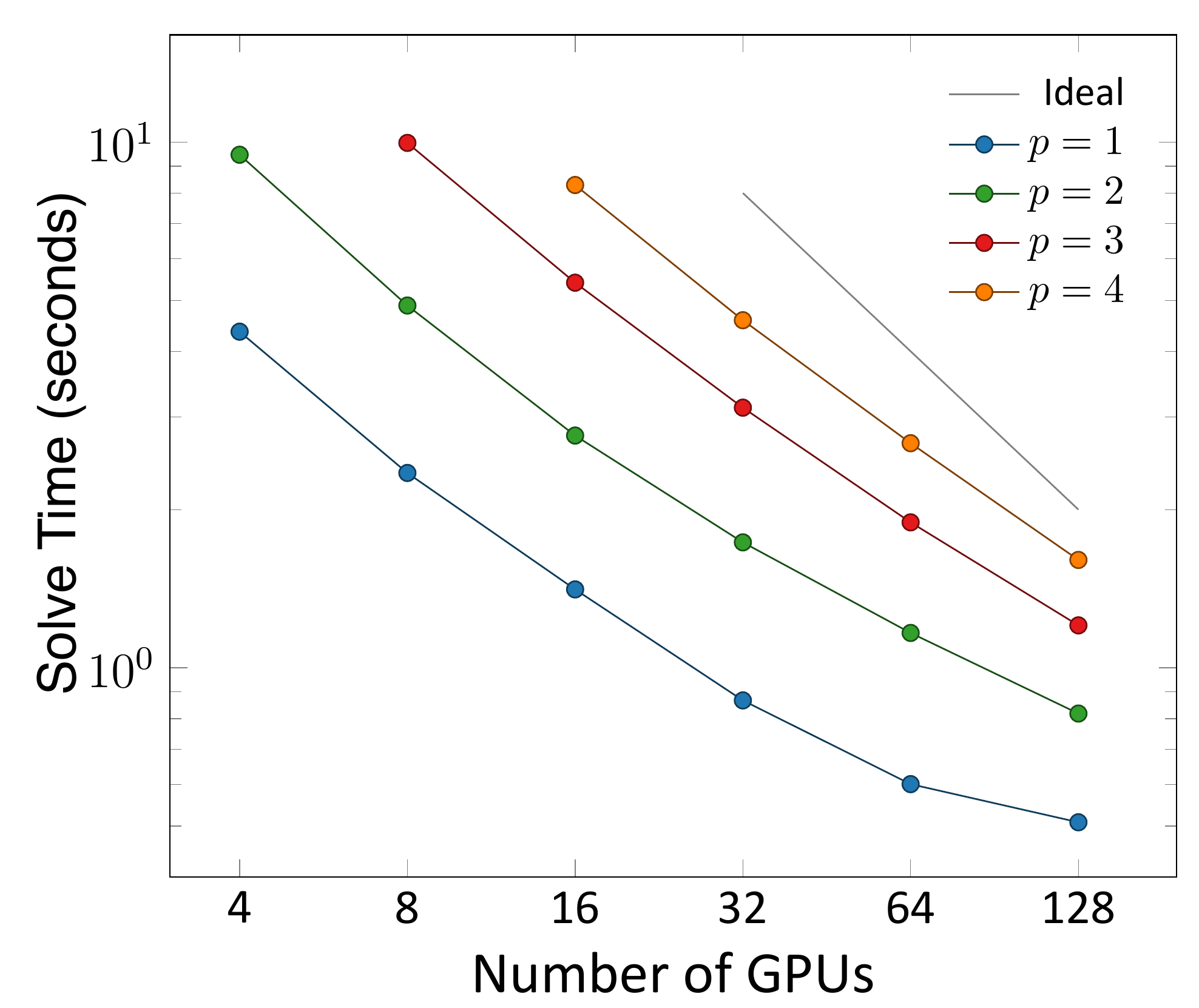}
\end{center}
\vspace{-7mm}
\caption{Strong scaling for the Kershaw benchmark.}
\label{fig_strong}
\end{figure}

A strong scaling study of the GPU algorithm is reported in
Figure \ref{fig_strong} on up to 128 GPUs.
For all computations, the number of elements is fixed to
$96 \times 96 \times 96$.
For ideal strong scaling, the time per cycle would decrease linearly as we
increase the GPU count.
As expected, the scaling deteriorates as the GPU count increases,
as there is no longer sufficient work to feed each GPU.

Every data point in Figure \ref{fig_strong} is obtained by timing the
computation of a single Newton iteration on the initial deformed mesh,
with 20 iterations of the MINRES linear solver.
The number of quadrature point per element is set to $3^3, 4^3, 5^3$,
and $6^3$ for mesh orders $1,2,3$, and $4$, respectively.

\paragraph{Throughput}

\begin{figure}[htbp!]
\begin{center}
$\begin{array}{cc}
\includegraphics[height=0.45\textwidth]{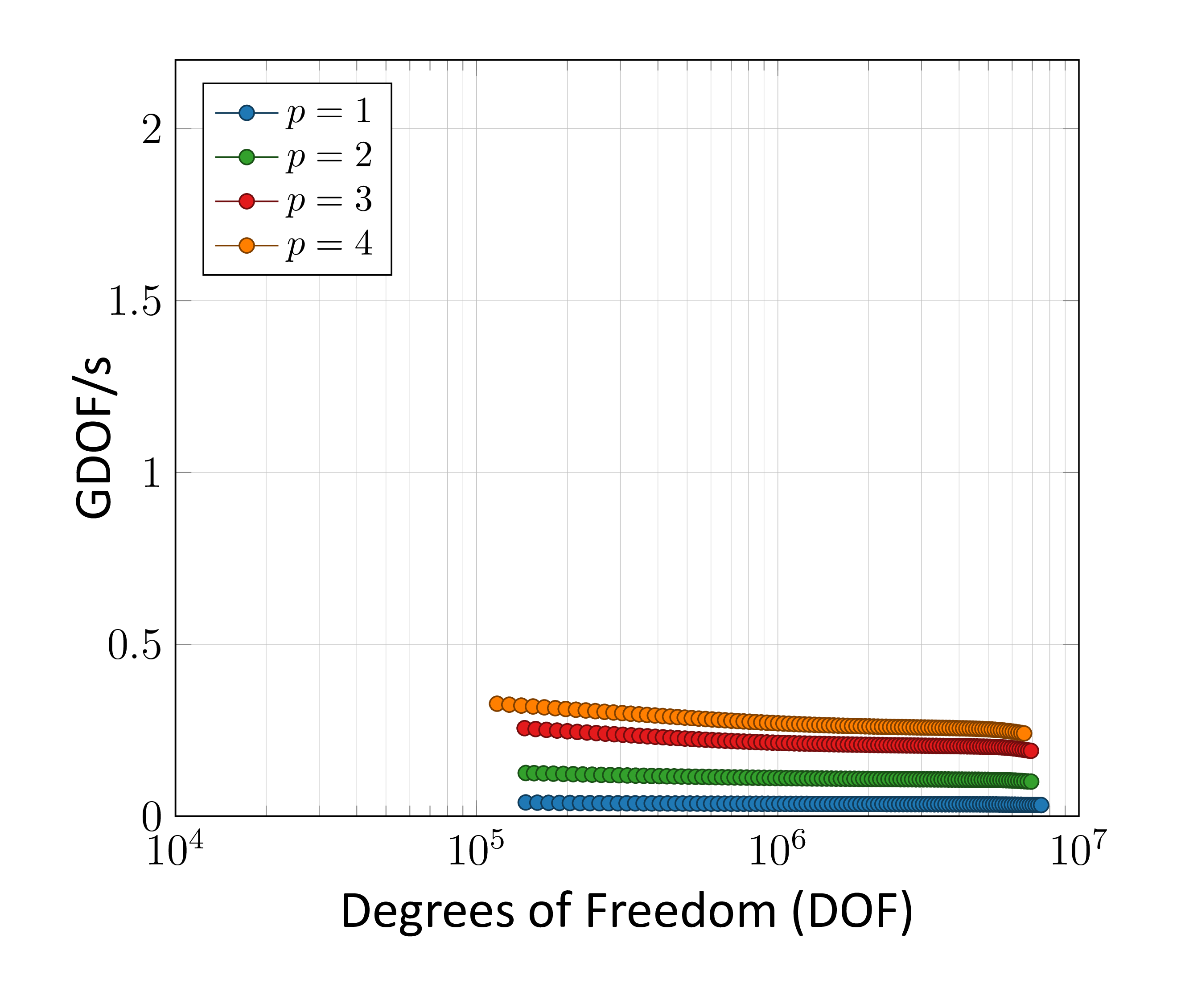} &
\includegraphics[height=0.45\textwidth]{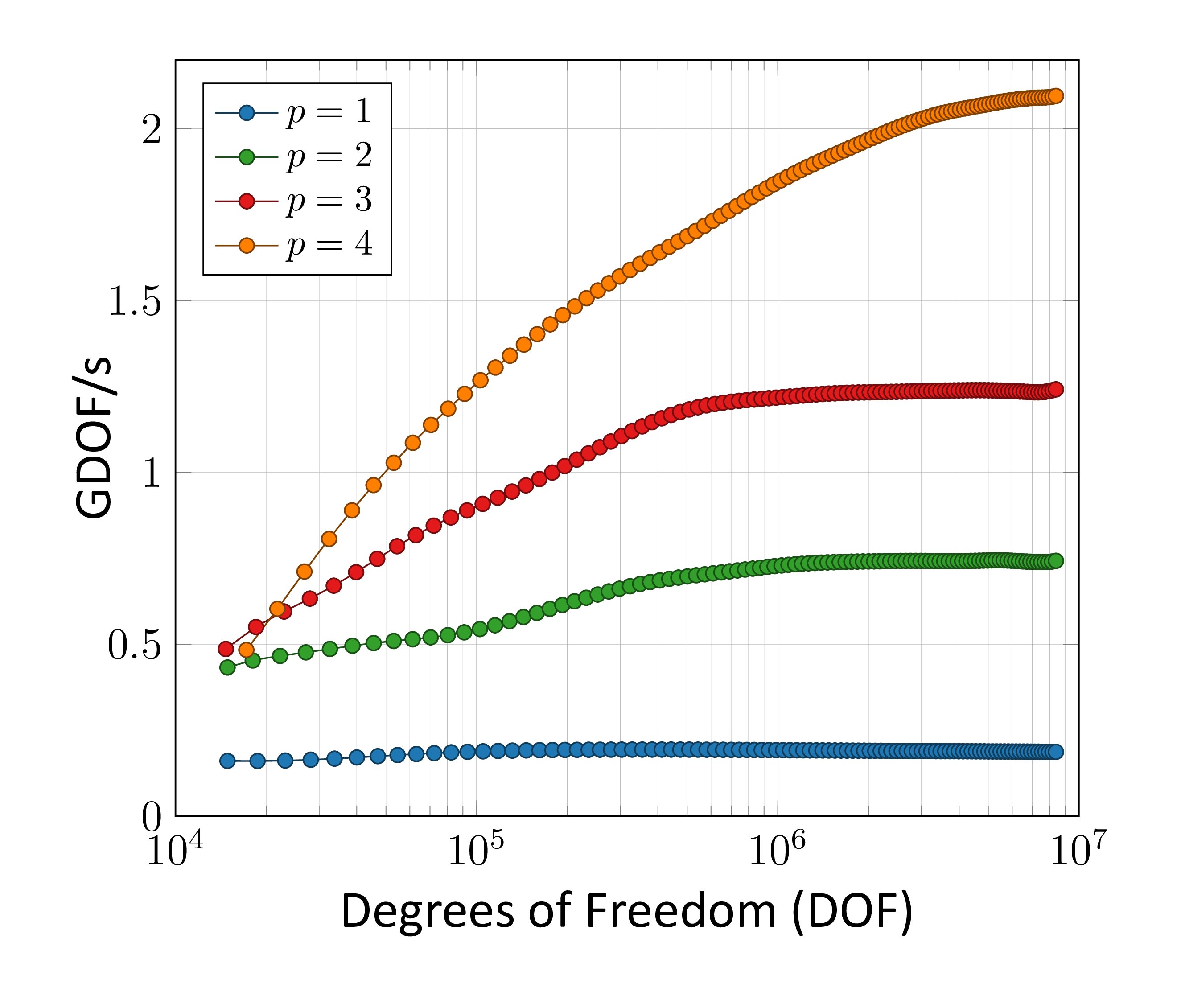} \\[-2em]
\mbox{a. 40 IBM Power9 cores}  & \mbox{b. 4 NVIDIA Tesla V100-SXM2} \\
\end{array}$
\end{center}
\vspace{-7mm}
\caption{Throughput comparison - action of the second derivative operator.
         Single Lassen node throughput
        (in GDOF/s, i.e. billions of degrees of freedom per second) of
        (a) 40 IBM Power9 CPU cores and
        (b) 4 NVIDIA Tesla V100-SXM GPU.}
\label{fig_kernel_tput}
\end{figure}

Next we provide throughput results that demonstrate how well the proposed
algorithms utilize the machine resources as the problem size increases.
The plots on Figure \ref{fig_kernel_tput} show the throughput
for our most important kernel, i.e., the action of $\partial^2 F$,
in terms of billions of degrees of freedom processed per second vs.
the problem size in the CPU and GPU cases.
Such plots are useful in comparing the performance of different orders and
problem size
in both the weak and strong scaling limits, see e.g. \cite{CEED2021}.
For example, higher throughput means faster run time, and from Figure
\ref{fig_kernel_tput} we can see that on both platforms higher orders perform
better, but the difference is much more significant on GPUs.
Furthermore, while CPU performance is relatively flat across problem sizes, the
GPU requires significant number of unknowns (in the millions of degrees of
freedom) to achieve the best results.

\begin{figure}[h!]
\begin{center}
\includegraphics[height=0.5\textwidth]{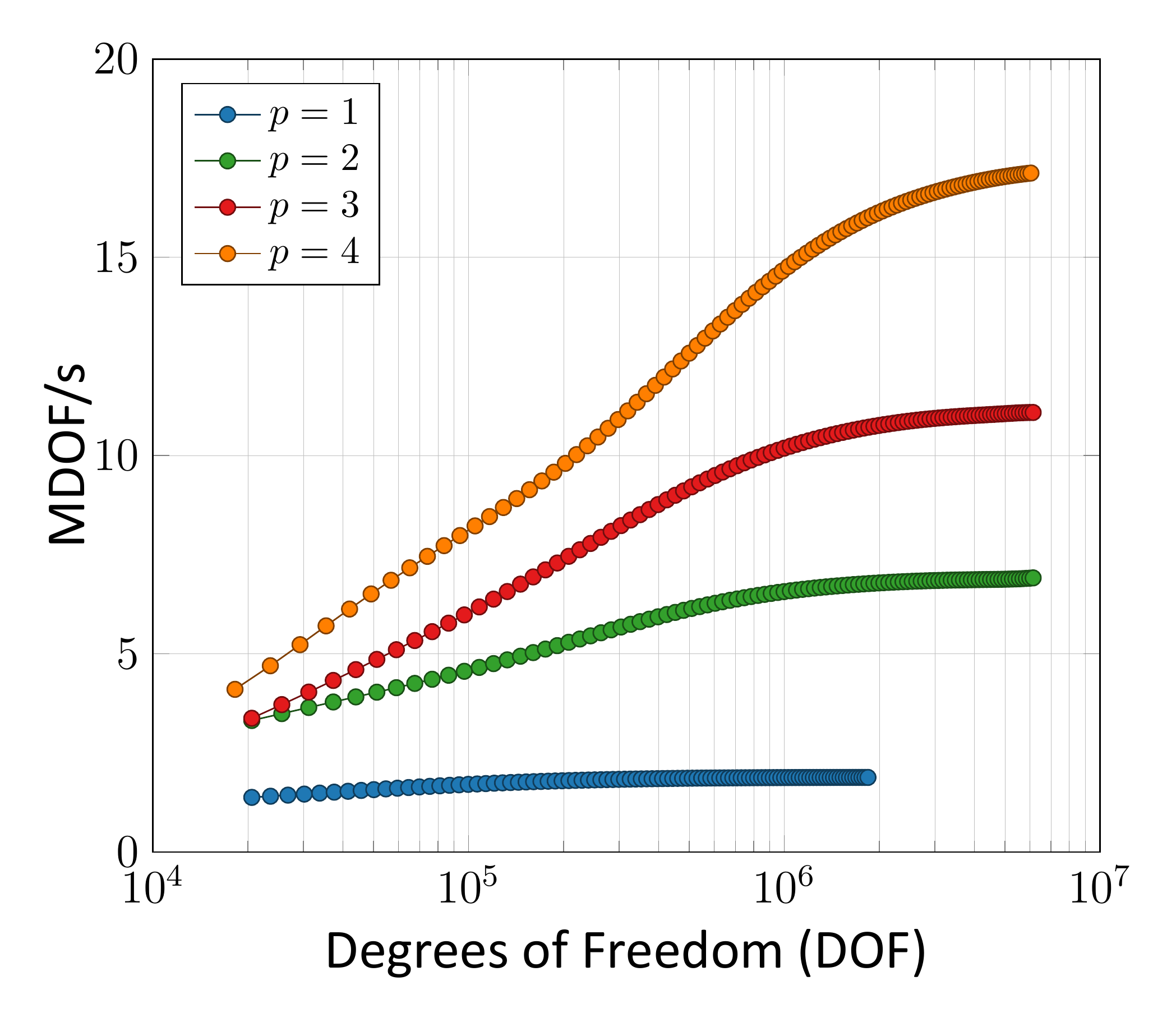}
\end{center}
\vspace{-7mm}
\caption{Throughput of the complete TMOP computation.
         Single NVIDIA Tesla V100-SXM GPU throughput
         (in MDOF/s, i.e. millions of processed degrees of freedom per second).}
\label{fig_full_tput}
\end{figure}

Figure \ref{fig_full_tput} shows the throughput for the complete TMOP algorithm
for a single GPU. Again we observe that the higher orders achieve better
computational intensity, especially for larger problems.
Every data point in Figures \ref{fig_kernel_tput}
and \ref{fig_full_tput} is obtained by timing the computation of a
single Newton iteration on the initial deformed mesh.
The number of linear solver iterations is fixed to 20 to make sure
every data point represents the same amount of computational work.
The number of quadrature points per element is set
to $3^3, 4^3, 5^3$, and $6^3$ for mesh orders $1,2,3$, and $4$, respectively.

\subsection{ALE Simulation with Material-Adaptive TMOP}
\label{subsec_ale}
In this section we demonstrate the proposed GPU-algorithms in the settings of a
multimaterial ALE hydrodynamics production code.
We consider a simulation of the BRL81a shaped charge using the MARBL
multiphysics application \cite{marbl}.
A shaped charge is a device which focuses the pressures of a high explosive onto
a metal ``liner'' to form a hyper-velocity jet which can be used in many
applications, including armor penetration, metal cutting and perforation of
wells for the oil/gas industry~\cite{walters2008brief}.


The ALE hydrodynamics algorithm consists of the following phases:
\begin{enumerate}
\item High-order Lagrange multimaterial hydrodynamics on a moving, unstructured,
  high-order mesh~\cite{dobrev2012high}, including use of GPU
  accelerated 3rd party libraries for material equations of state (EOS)
  evaluation and material constitutive models.
\item Non-linear, material adaptive, high-order mesh optimization using the
  presented TMOP method.
\item High-order continuous (kinematic) and discontinuous Galerkin
  (thermodynamic) advection-based remap using flux corrected transport (FCT).
\end{enumerate}

\begin{figure}[h!]
\begin{center}
$\begin{array}{cccc}
\includegraphics[height=0.22\textwidth]{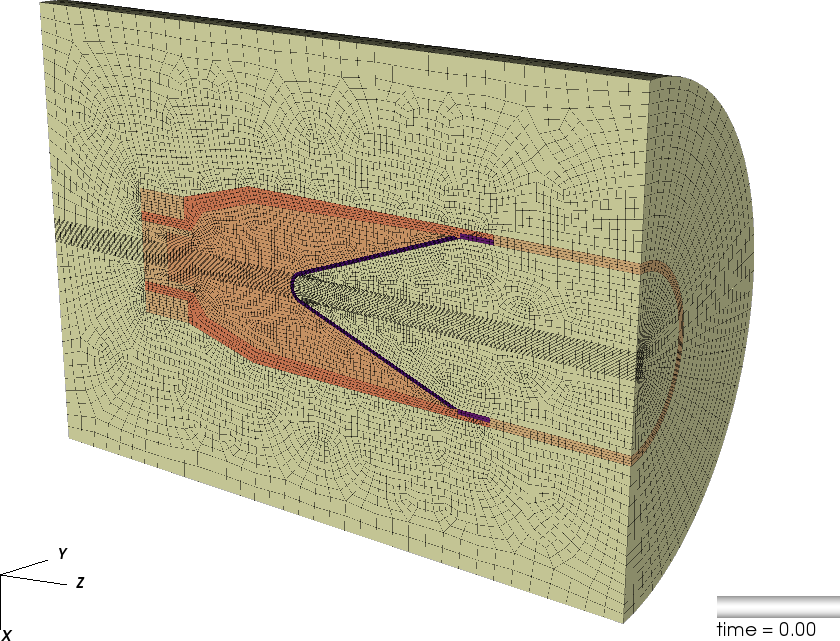} &
\includegraphics[height=0.22\textwidth]{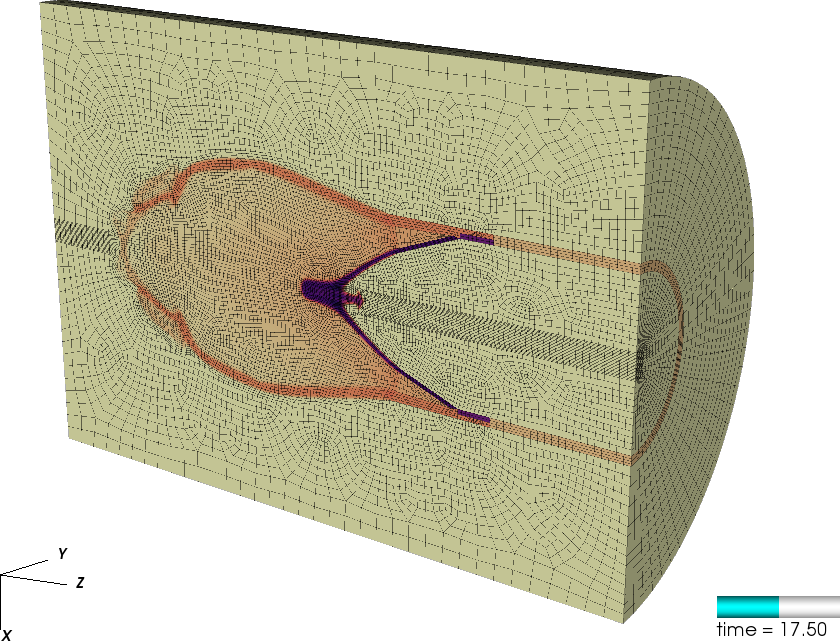} &
\includegraphics[height=0.22\textwidth]{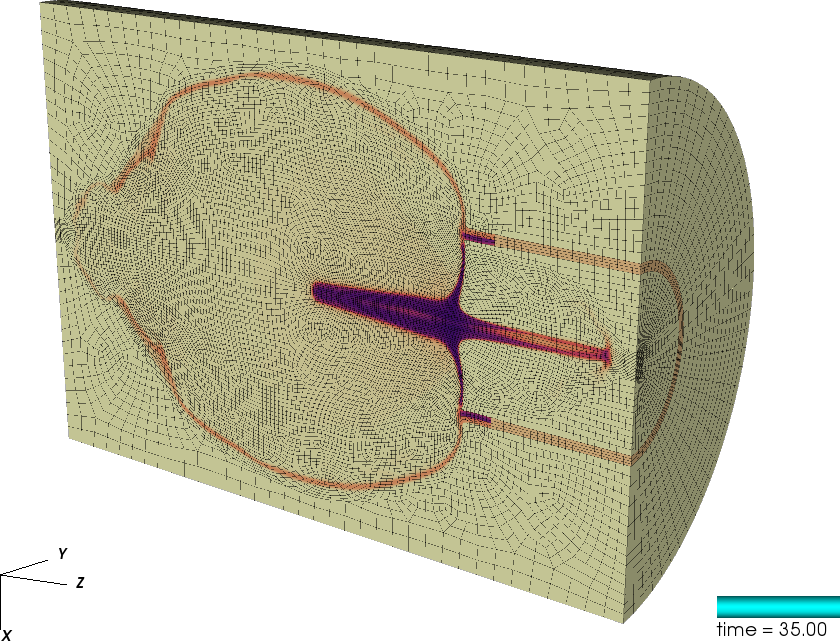} \\
\includegraphics[height=0.22\textwidth]{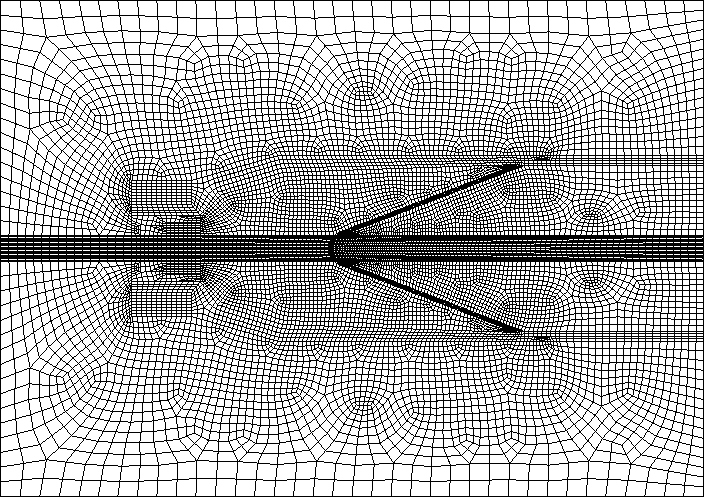} &
\includegraphics[height=0.22\textwidth]{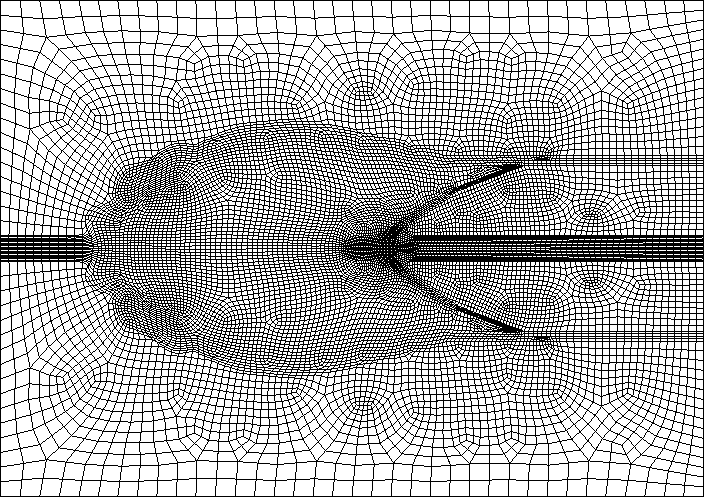} &
\includegraphics[height=0.22\textwidth]{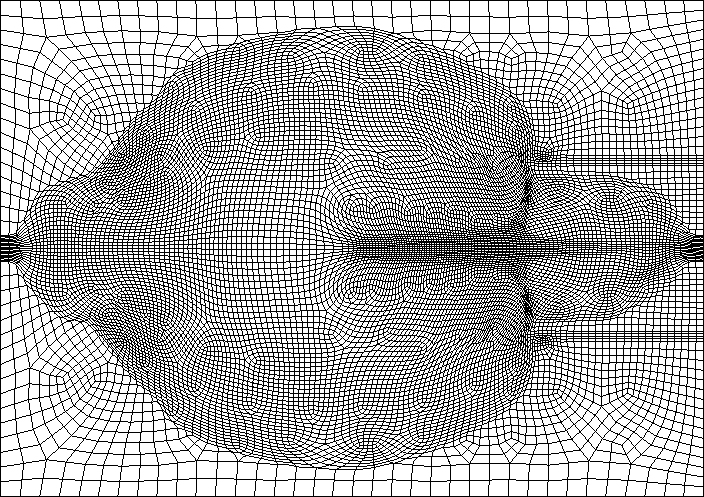} \\
\end{array}$
\end{center}
\vspace{-7mm}
\caption{Snapshots of the density (\emph{top}) and mesh (\emph{bottom}) for the ALE shaped
charge GPU simulation using material adaptive TMOP at times t=0, 17.5 and 35 (\emph{left to right}).}
\label{fig_shaped_charge}
\end{figure}

In Figure~\ref{fig_shaped_charge}, we show results of a MARBL calculation of the
BRL81a model with all three phases of the ALE algorithm run on the GPU.
To enhance the high-order mesh resolution near the hyper-velocity copper jet, we
employ the material adaptive capabilities of the TMOP mesh optimization phase at
the copper material with a 2:1 size ratio.
To prevent the mesh from moving in regions which have not experienced
hydrodynamic motion, we utilize the limiting term in our optimization metric.
We use $p=2$ elements for the mesh
and the continuous kinematic fields with $64$ quad points per element
(combined with $p=1$ discontinuous finite elements
for the thermodynamic variables).
This simulation consists of $14,346,240$ total quadrature points on a
high-order ($p=2$), highly unstructured mesh.
We run the problem using 2 nodes of the Lassen machine at LLNL and compare
performance results between the CPU only case (80 IBM Power9 cores) and the GPU
case (8 NVIDIA Tesla V100-SXM GPUs) for total of 2500 simulation cycles.
Both cases utilize the partial assembly approach.
Over these 2500 cycles, the code performs ALE remap at fixed time intervals for
a total of 25 ALE steps.
The material adaptive TMOP Newton solver is run at each ALE step with an outer
residual tolerance of $10^{-10}$ with a max of 10 iterations while the inner
linear solver has a max of $100$ iterations.
The relative performance comparison between the CPU and GPU node simulations is
shown in Table~\ref{tab_shaped_charge_1} and Table~\ref{tab_shaped_charge_2}.
In the CPU case, mesh optimization takes a substantially larger portion of the
overall runtime, comprising $15\%$ of total wall time vs. $7\%$ in the GPU case
as shown in Table~\ref{tab_shaped_charge_1}.
The GPU speedup for the mesh optimization phase of the simulation is over
$20\times$ as shown in Table~\ref{tab_shaped_charge_2}.

\begin{table}[h!]
\begin{center}
\begin{tabular}{| c | c | c | c | c | }
\hline
    & Lagrange & Mesh Optimization & Remap   & Other   \\
\hline
CPU\textsuperscript{PA} &  31.34\% & 14.86\%       & 53.53\% & 0.17\%  \\
\hline
GPU\textsuperscript{PA} &  14.00\% & 6.72\%        & 73.39\% & 5.89\%  \\
\hline
\end{tabular}
\caption{Percentage of total run time for each phase of the multimaterial ALE simulation. Comparison of CPU vs GPU runs.}
\label{tab_shaped_charge_1}
\end{center}
\end{table}

\begin{table}[h!]
\begin{center}
\begin{tabular}{| l | r | r | r | r |}
\hline
    & Min time/rank & Max time/rank & Avg time/rank  & Speedup \\
\hline
CPU\textsuperscript{PA} MeshOpt & 4730.828   & 4730.831 & 4730.830 & - \\
CPU\textsuperscript{PA} Action  & 4713.425   & 4713.427 & 4713.426 & - \\
\hline
GPU\textsuperscript{PA} MeshOpt & 288.126    & 289.125  & 288.842  &
\textbf{16.37$\times$} \\
GPU\textsuperscript{PA} Action  & 209.422    & 209.436  & 209.430  &
\textbf{22.51$\times$} \\
\hline
\end{tabular}
\caption{Total run time with min/max/avg time across all MPI ranks
        (80 for CPU, 8 for GPU) for the full mesh optimization phase (MeshOpt)
        and for the $\partial^2 F$ operator evaluation (Action) over 2500
        simulation cycles for the multimaterial ALE simulation.
        Comparison of CPU vs GPU runs.}
\label{tab_shaped_charge_2}
\end{center}
\end{table}


\section{Conclusion}
This paper discussed the use of finite element partial assembly techniques in
the context of high-order mesh optimization.
We demonstrated that this approach leads to substantial improvements in
performance complexity for tensor-product elements (quads and hexes), and the
resulting tensor contractions perform well on GPU hardware.
In addition, we proposed a simple mesh optimization benchmark for the mesh
optimization community.

Many of the approaches in this paper do extend to simplicial meshes,
and we plan to address that in our future work.
Specifically, the finite element operator decomposition \eqref{eq_decompose},
the concept of partial assembly, and all computations at quadrature points
do extend to simplices.
However, simplices do not support a tensor-product
structure \eqref{eqn_basis_tensor} and one can't use sum factorization
to transition from degrees of freedom to quadrature points.
Our method also relies on efficient matrix-free preconditioning, which is
still an area of active research.
We plan to explore that further in a future publication as well.


\bibliographystyle{elsarticle-num}
\bibliography{gpu}


\appendix

\newpage

\section{Source Code for Generating the Kershaw Mesh}
\label{sec_appendix}
In this appendix we provide the C++ code that is used to obtain the initial 3D
Kershaw mesh that is optimized in the benchmark tests of Section
\ref{subsec_kershaw}.
The initial domain is $[0,1]^3$.
Starting with a Cartesian mesh, the deformed configuration is computed
by the \code{kershaw} function that transforms the input coordinates
\code{x,y,z} to the deformed positions \code{X,Y,Z}.
The deformation is applied to all nodes of the position function $x$,
see Section \ref{subsec_mesh}.
If the initial Cartesian mesh is $n_x \times n_y \times n_z$, then $n_x$
should be divisible by 6 and $n_y$ and $n_z$ should be divisible by 2.
The parameters \code{epsy} and \code{epsz} correspond to $\epsy$ and $\epsz$,
see Figure \ref{fig_kershaw_meshes}.

\medskip

\begin{lstlisting}
double right(const double eps, const double x) // 1D transformation at right boundary
{
   return (x <= 0.5) ? (2-eps)*x : 1+eps*(x-1);
}
double left(const double eps, const double x)  // 1D transformation at left boundary
{
   return 1-right(eps,1-x);
}
double step(const double a, const double b, double x)
{
   if (x <= 0) { return a; }
   if (x >= 1) { return b; }
   return a + (b-a)*(x*x*x*(x*(6*x-15)+10));   // Smooth transition from a to b
}
void kershaw(const double epsy, const double epsz,
             const double x, const double y, const double z,
             double &X, double &Y, double &Z)  // (x,y,z) -> (X,Y,Z) Kershaw transform
{
   X = x;
   int layer = x*6.0;
   double lambda = (x-layer/6.0)*6;
   switch (layer)
   {
      case 0:
         Y = left(epsy, y);
         Z = left(epsz, z);
         break;
      case 1:
      case 4:
         Y = step(left(epsy, y), right(epsy, y), lambda);
         Z = step(left(epsz, z), right(epsz, z), lambda);
         break;
      case 2:
         Y = step(right(epsy, y), left(epsy, y), lambda/2);
         Z = step(right(epsz, z), left(epsz, z), lambda/2);
         break;
      case 3:
         Y = step(right(epsy, y), left(epsy, y), (1+lambda)/2);
         Z = step(right(epsz, z), left(epsz, z), (1+lambda)/2);
         break;
      default:
         Y = right(epsy, y);
         Z = right(epsz, z);
         break;
   }
}
\end{lstlisting}

\end{document}